\documentclass[twocolumn,aps,pra,superscriptaddress,floatfix]{revtex4-2}
\usepackage{graphicx}
\usepackage{times}
\usepackage{nicefrac}
\usepackage{amsmath}
\usepackage{amsfonts}
\usepackage{amssymb}
\usepackage{amsthm}
\usepackage{epsf}
\usepackage{bm}
\usepackage{bbm}
\usepackage{times}

\usepackage{dcolumn}

\newcolumntype{.}{D{x}{}{-1}}
\newcolumntype{w}[1]{D{.}{.}{#1}}

\begin{document}

\newcommand{\half}{\frac12}
\newcommand{\vare}{\varepsilon}
\newcommand{\eps}{\epsilon}
\newcommand{\pr}{^{\prime}}
\newcommand{\ppr}{^{\prime\prime}}
\newcommand{\pp}{{p^{\prime}}}
\newcommand{\ppp}{{p^{\prime\prime}}}
\newcommand{\hp}{\hat{\bfp}}
\newcommand{\hr}{\hat{\bfr}}
\newcommand{\hk}{\hat{\bfk}}
\newcommand{\hx}{\hat{\bfx}}
\newcommand{\hpp}{\hat{\bfpp}}
\newcommand{\hq}{\hat{\bfq}}
\newcommand{\rqq}{{\rm q}}
\newcommand{\bfk}{{\bm{k}}}
\newcommand{\bfp}{{\bm{p}}}
\newcommand{\bfq}{{\bm{q}}}
\newcommand{\bfr}{{\bm{r}}}
\newcommand{\bfx}{{\bm{x}}}
\newcommand{\bfy}{{\bm{y}}}
\newcommand{\bfz}{{\bm{z}}}
\newcommand{\bfpp}{{\bm{\pp}}}
\newcommand{\bfppp}{{\bm{\ppp}}}
\newcommand{\balpha}{\bm{\alpha}}
\newcommand{\bvare}{\bm{\vare}}
\newcommand{\bgamma}{\bm{\gamma}}
\newcommand{\bGamma}{\bm{\Gamma}}
\newcommand{\bLambda}{\bm{\Lambda}}
\newcommand{\bmu}{\bm{\mu}}
\newcommand{\bnabla}{\bm{\nabla}}
\newcommand{\bvarrho}{\bm{\varrho}}
\newcommand{\bsigma}{\bm{\sigma}}
\newcommand{\bTheta}{\bm{\Theta}}
\newcommand{\bphi}{\bm{\phi}}
\newcommand{\bomega}{\bm{\omega}}
\newcommand{\intzo}{\int_0^1}
\newcommand{\intinf}{\int^{\infty}_{-\infty}}
\newcommand{\lbr}{\langle}
\newcommand{\rbr}{\rangle}
\newcommand{\ThreeJ}[6]{
        \left(
        \begin{array}{ccc}
        #1  & #2  & #3 \\
        #4  & #5  & #6 \\
        \end{array}
        \right)
        }
\newcommand{\SixJ}[6]{
        \left\{
        \begin{array}{ccc}
        #1  & #2  & #3 \\
        #4  & #5  & #6 \\
        \end{array}
        \right\}
        }
\newcommand{\NineJ}[9]{
        \left\{
        \begin{array}{ccc}
        #1  & #2  & #3 \\
        #4  & #5  & #6 \\
        #7  & #8  & #9 \\
        \end{array}
        \right\}
        }
\newcommand{\Vector}[2]{
        \left(
        \begin{array}{c}
        #1     \\
        #2     \\
        \end{array}
        \right)
        }

\newcommand{\Dmatrix}[4]{
        \left(
        \begin{array}{cc}
        #1  & #2   \\
        #3  & #4   \\
        \end{array}
        \right)
        }
\newcommand{\Dcase}[4]{
        \left\{
        \begin{array}{cl}
        #1  & #2   \\
        #3  & #4   \\
        \end{array}
        \right.
        }
\newcommand{\cross}[1]{#1\!\!\!/}

\newcommand{\Za}{{Z \alpha}}
\newcommand{\im}{{ i}}

\title{QED calculations of energy levels of helium-like ions with $\bm {5 \leq Z \leq 30}$ }

\author{Vladimir A. Yerokhin}
\affiliation{Peter the Great St.~Petersburg Polytechnic University,
Polytekhnicheskaya 29, 195251 St.~Petersburg, Russia}

\author{Vojt\v{e}ch Patk\'o\v{s}}
\affiliation{Faculty of Mathematics and Physics, Charles University,  Ke Karlovu 3, 121 16 Prague 2, Czech Republic}

\author{Krzysztof Pachucki}
\affiliation{Faculty of Physics, University of Warsaw,
             Pasteura 5, 02-093 Warsaw, Poland}

\date{\today}

\begin{abstract}

A calculation of two-electron QED effects to all orders in the nuclear binding strength parameter
$\Za$ is presented for the ground and $n = 2$ excited states of helium-like ions. After subtracting
the first terms of the $\Za$ expansion from the all-order results, we identify the higher-order
QED effects of order $m\alpha^7$ and higher. Combining the higher-order remainder with the
results complete through order $m\alpha^6$  from [V.~A.~Yerokhin and K.~Pachucki, Phys.~Rev.~A {\bf 81},
022507 (2010)], we obtain the most accurate theoretical predictions for the ground and non-mixing
$n=2$ states of helium-like ions with $Z = 5\,$--$\,30$. For the mixing $2\,^1 P_1$ and $2\,^3 P_1$ states,
we extend the previous calculation by evaluating
the higher-order mixing correction and show that it defines the uncertainty
of theoretical calculations in the $LS$ coupling for $Z > 10$.

\end{abstract}

\maketitle

\section{Introduction}

Helium and helium-like ions are the simplest few-body atomic systems in nature. They have been extensively
studied experimentally, measurements being facilitated by the availability of
long-lived metastable $^3S$ and $^3P$ states in their spectrum.  The simplicity and the long
history of experimental studies make helium and
helium-like ions a widely used testing ground for different methods of
atomic-structure calculations. Alongside with the hydrogen-like atoms, the helium-like ions are the systems
for which theoretical predictions are the most accurate \cite{indelicato:19}.

There are two essentially different approaches in {\em ab initio} QED calculations of atomic energies.
The first method, often referred to as the ``all-order'' approach, starts with the Dirac energies and accounts
for the electron-electron interaction by a systematic QED perturbation expansion. In this way one includes
all orders in the electron-nucleus binding strength parameter $Z\alpha$ (where $Z$ is the nuclear charge
number and $\alpha$ is the fine-structure constant) but expands in the electron-electron interaction parameter
$1/Z$. This method yields very accurate results for high-$Z$ ions but its accuracy gradually diminishes
as $Z$ decreases because of uncertainties associated with higher-order
electron-correlation effects. Calculations by this method were carried
out for helium-like ions with $Z \ge 12$ in Refs.~\cite{artemyev:05:pra,kozhedub:19,malyshev:19}.

For low-$Z$ systems and especially for the helium atom, the best results are obtained with the method
of the nonrelativistic QED (NRQED). The starting point is the few-body
Schr\"odinger equation, which includes the Coulomb electron-electron interaction to all orders. The
relativistic and QED effects are accounted for perturbatively, with the expansion parameters $\alpha$ and $Z\alpha$.
The NRQED approach yields very accurate results for the helium atom \cite{pachucki:17:heSummary,yerokhin:21:hereview}
but the uncertainties due to higher-order QED terms are enhanced by high powers of $Z$ and thus grow fast
for helium-like ions as $Z$ increases.

In order to achieve more accurate results, pure NRQED calculations need to be
complemented by a separate treatment of higher-order (in $\Za$) effects, based on the all-order calculations.
In this way, the two complementary approaches can be merged, yielding an improved accuracy in the region of medium
$Z$, where each of these methods separately encounter difficulties. This ``unified'' approach was introduced
for helium-like ions by Drake in Ref.~\cite{drake:88:cjp}. It was advanced further
in our previous investigation \cite{yerokhin:10:helike}, where all
effects up to order $m\alpha^6$ were taken into account. Recently, the unified approach was applied to
obtain the most accurate predictions for the fine structure of lithium-like ions \cite{yerokhin:20:fs}.

The limitation on the accuracy of our previous work on helium-like ions \cite{yerokhin:10:helike} was
coming from the fact that
the higher-order QED effects were obtained with the inclusion
of one-electron effects only (i.e., to the zeroth order in $1/Z$). The identification of the higher-order
two-electron QED contributions was not possible at that time, because of insufficient numerical accuracy of
the all-order QED calculations.

The goal of the present investigation is to perform a high-precision numerical calculation of the two-electron
QED corrections to all orders in $\Za$ and to identify the higher-order $m\alpha^{7+}$ QED contribution to first
order in $1/Z$. The higher-order correction obtained in this way is then
added to the results of Ref.~\cite{yerokhin:10:helike},
yielding improved theoretical predictions for ionization energies of helium-like ions.

An additional motivation for the present investigation comes from the observation of a
significant discrepancy between
theoretical predictions and experimental results for the ionization energies of the triplet $n = 2$ states
in the helium atom \cite{patkos:21:helamb,clausen:21}. In view of this discrepancy, it is important to
cross-check the NRQED evaluation of higher-order QED effects
\cite{patkos:21:helamb,yerokhin:18:betherel,patkos:20} against the all-order calculations.

The paper is constructed as follows. Sec.~\ref{sec:notation} presents a set of definitions and
notations to be extensively used throughout the paper. In Secs.~\ref{sec:se}, \ref{sec:vp}, and \ref{sec:2ph} we
describe our  all-order calculations of the screened self-energy, the screened vacuum-polarization, and the
two-photon exchange corrections, respectively. In Sec.~\ref{sec:nonmix} we compare
the all-order results for the non-mixing states with calculations performed within the
$Z\alpha$-expansion approach and identify the remainder of order $m\alpha^7$ and higher.
In Sec.~\ref{sec:mix} we describe the comparison of the two approaches for the mixing
$2\,^1 P_1$ and $2\,^3 P_1$ states and calculate the mixing correction
of order $m\alpha^7$ and higher. Finally, Sec.~\ref{sec:results} presents our results for the
ionization energies and compares theoretical predictions with experimental data.
The relativistic units $\hbar=c=1$ and Heaviside charge units $\alpha = e^2/(4\pi)$
are used in the paper, unless explicitly specified otherwise.

\section{Definitions}
\label{sec:notation}

Throughout the paper, we will extensively use several basic QED operators:
the operator of the
electron-electron interaction $I(\omega)$, the one-loop self-energy operator $\Sigma(\vare)$,
and the one-loop vacuum-polarization potential $U_{\rm VP}$.
The operator of the
electron-electron interaction $I(\omega)$ is defined by
\begin{equation}\label{a1}
  I(\omega,\bfr_{1},\bfr_{2}) = e^2\, \alpha_{1}^{\mu} \alpha_{2}^{\nu}\, D_{\mu\nu}(\omega,\bfr_{12})\,,
\end{equation}
where $\alpha^{\mu} = (1,\balpha)$ are the Dirac matrices, $\bfr_{12} = \bfr_{1} - \bfr_{2}$, and
$D_{\mu\nu}(\omega,\bfr_{12})$ is the photon propagator.
The one-loop self-energy operator $ \Sigma(\vare)$ is
defined by its matrix elements with one-electron wave functions $|a\rbr$ and
$|b\rbr$ as
\begin{align}
\lbr a | \Sigma(\vare)| b\rbr = \frac{i}{2\pi}\intinf d\omega \sum_n
  \frac{\lbr an| I(\omega)| nb\rbr}{\vare-\omega - u\vare_n}\,,
\end{align}
where the sum over $n$ is extended over the complete spectrum of the Dirac equation (implying
the summation over the discrete states and the integration over the continuum part
of the spectrum) and $u = 1 -i\epsilon$ is the infinitesimal addition which ensures the correct
position of poles of the electron propagator with respect to the integration contour.
The vacuum-polarization potential $U_{\rm VP}$ is given by
\begin{align}
U_{\rm VP}(\bfx) = \frac{\alpha}{2\pi i}
 \intinf d\omega \, \int d^3\bfy \, \frac{1}{|\bfx-\bfy|}\,{\rm Tr} \big[ G(\omega,\bfy,\bfy)\big]\,,
\end{align}
where $G(\omega) = (\omega-h_D)^{-1}$ is the Dirac-Coulomb Green function and $h_D$ is the one-electron
Dirac-Coulomb Hamiltonian.

In order to simplify the following formulas,
we use the following short-hand notations \cite{artemyev:05:pra} for the summations
over the Clebsch-Gordan coefficients in the initial- and the final-state two-electron wave
function,
\begin{align}
F_i |i_1i_2\rbr \equiv N \sum_{\mu_{i_1}\mu_{i_2}} C_{j_{i_1}\mu_{i_1},j_{i_2}\mu_{i_2}}^{JM}\,
 |i_1i_2\rbr\,,
\end{align}
where $|i_1i_2\rbr$ is the direct product of the one-electron wave functions $|i_1i_2\rbr =
|j_{i_1}l_{i_1}\mu_{i_1}\rbr|j_{i_2}l_{i_2}\mu_{i_2}\rbr$ and $N$ is the normalization factor, $N
= 1$ for the non-equivalent electrons ($i_1\ne i_2$) and $N = 1/\sqrt{2}$ for the case of
equivalent electrons ($i_1 = i_2$). We also introduce the permutation operators $P$ and $Q$ that interchange
the initial-state and final-state electrons, respectively,
\begin{align}
\sum_{P}(-1)^P|Pi_1Pi_2\rbr \equiv |i_1i_2\rbr - |i_2i_1\rbr\,,\\
\sum_{Q}(-1)^Q|Qk_1Qk_2\rbr \equiv |k_1k_2\rbr - |k_2k_1\rbr\,,
\end{align}
where the summation indices $P$ and $Q$ indicates that the summation is carried out over all permutations
and $(-1)^P$ and $(-1)^Q$ are the signs of the permutations.

The formulas for the two-electron QED corrections in this paper will be written in the form that is
often non-symmetric with respect to the interchange of the initial- and the final-state electron states.
It is often the case that the symmetry is preserved and can be proved explicitly. The exception is the
case of non-diagonal matrix elements occuring for the mixing $^1P_1$ and $^3P_1$ states; in this
case it is assumed that the formulas should
be symmetrized with respect $(i_1i_2) \leftrightarrow (k_1k_2)$.

We will also use following the shorthand notations: $\Delta_{a,b} = \vare_a-\vare_b$,
$I'(\omega) = \partial I(\omega)/(\partial \omega)$, and $\Sigma{}'(\vare) =
\partial \Sigma(\vare)/(\partial \vare)$.
\\

\section{Screened self-energy}
\label{sec:se}

\begin{table*}[t]
\caption{The screened
self-energy correction for the $n = 1$ and $n = 2$ states of He-like ions, in units of
$\alpha^2(\Za)^3$.
\label{tab:sescr}}
\begin{ruledtabular}
\scriptsize
\begin{tabular}{lw{3.8}w{3.8}w{3.8}w{3.8}w{3.8}w{3.8}w{3.8}w{2.10}l}
\multicolumn{1}{l}{$Z$}
& \multicolumn{1}{c}{$(1s1s)_0$}
  & \multicolumn{1}{c}{$(1s2s)_0$}
    & \multicolumn{1}{c}{$(1s2s)_1$}
        & \multicolumn{1}{c}{$(1s2p_{1/2})_0$}
            & \multicolumn{1}{c}{$(1s2p_{1/2})_1$}
                & \multicolumn{1}{c}{$(1s2p_{3/2})_1$}
                    & \multicolumn{1}{c}{$(1s2p_{3/2})_2$}
                        & \multicolumn{1}{c}{offdiag}
                        & \multicolumn{1}{c}{}
  \\\hline\\[-5pt]

 10  & -2.410\,58\,(36)& -0.524\,90\,(14)& -0.329\,31\,(11)& -0.101\,63\,(5) & -0.075\,82\,(11)& -0.059\,00\,(11)& -0.145\,74\,(12)& 0.058\,372\,(12)\\
 12  & -2.213\,84\,(15)& -0.484\,46\,(16)& -0.303\,28\,(5) & -0.091\,99\,(3) & -0.069\,47\,(9) & -0.056\,00\,(9) & -0.135\,26\,(9) & 0.053\,223\,(7)\\
 14  & -2.054\,26\,(9) & -0.451\,78\,(10)& -0.282\,14\,(3) & -0.084\,39\,(4) & -0.064\,45\,(9) & -0.053\,61\,(9) & -0.126\,75\,(7) & 0.049\,049\,(6)\\
 16  & -1.921\,71\,(9) & -0.424\,79\,(9) & -0.264\,54\,(3) & -0.078\,30\,(4) & -0.060\,43\,(8) & -0.051\,69\,(8) & -0.119\,69\,(7) & 0.045\,582\,(3)\\
 18  & -1.809\,65\,(8) & -0.402\,13\,(7) & -0.249\,63\,(4) & -0.073\,36\,(4) & -0.057\,15\,(6) & -0.050\,12\,(6) & -0.113\,73\,(7) & 0.042\,648\,(2)\\[3pt]
 20  & -1.713\,71\,(6) & -0.382\,86\,(3) & -0.236\,82\,(2) & -0.069\,32\,(2) & -0.054\,46\,(3) & -0.048\,82\,(4) & -0.108\,63\,(4) & 0.040\,129\,(2)\\
     & -1.713\,72\,(8) & -0.382\,86\,(3) & -0.236\,82\,(3) & -0.069\,32\,(3) & -0.054\,46\,(3) & -0.048\,83\,(3) & -0.108\,63\,(3) & 0.040\,129\,(1)&Coulomb\\
     & -1.713\,7\,(3)  & -0.382\,8\,(3)  & -0.236\,8\,(3)  & -0.069\,3\,(1)  & -0.054\,4\,(1)  & -0.048\,8\,(2)  & -0.108\,6\,(2)  & 0.040\,13\,(3) & Ref.~\cite{artemyev:05:pra}\\
     & -1.713\,4\,(3)  &                 &                 &                 &                 &                 &                 &              & Ref.~\cite{yerokhin:97:pla}\\[3pt]
 24  & -1.558\,27\,(6) & -0.352\,04\,(4) & -0.215\,97\,(2) & -0.063\,33\,(2) & -0.050\,46\,(3) & -0.046\,88\,(3) & -0.100\,35\,(5) & 0.036\,020\,(2)\\
 28  & -1.438\,56\,(7) & -0.328\,85\,(3) & -0.199\,79\,(1) & -0.059\,41\,(1) & -0.047\,83\,(2) & -0.045\,58\,(3) & -0.093\,96\,(4) & 0.032\,811\,(3)\\[3pt]
 30  & -1.388\,86\,(4) & -0.319\,43\,(1) & -0.193\,02\,(1) & -0.058\,06\,(1) & -0.046\,93\,(2) & -0.045\,12\,(3) & -0.091\,31\,(4) & 0.031\,457\,(1)\\
     & -1.388\,8\,(2)  & -0.319\,4\,(2)  & -0.193\,0\,(2)  & -0.058\,1\,(1)  & -0.047\,0\,(1)  & -0.045\,2\,(2)  & -0.091\,3\,(2)  & 0.031\,46\,(2)& Ref.~\cite{artemyev:05:pra}\\
     & -1.388\,9\,(3)  &                 &                 &                 &                 &                 &                 &              & Ref.~\cite{yerokhin:97:pla}\\[3pt]
 32  & -1.344\,72\,(3) & -0.311\,22\,(2) & -0.186\,99\,(1) & -0.057\,05\,(1) & -0.046\,25\,(1) & -0.044\,77\,(2) & -0.088\,94\,(4) & 0.030\,239\,(1)\\
 36  & -1.270\,43\,(3) & -0.297\,88\,(1) & -0.176\,77\,(1) & -0.055\,95\,(1) & -0.045\,49\,(1) & -0.044\,33\,(2) & -0.084\,94\,(4) & 0.028\,138\,(1)\\
 40  & -1.211\,48\,(1) & -0.287\,96\,(1) & -0.168\,57\,(1) & -0.055\,89\,(1) & -0.045\,42\,(1) & -0.044\,19\,(2) & -0.081\,72\,(2) & 0.026\,393\,(1)\\
\end{tabular}
\end{ruledtabular}
\end{table*}

The derivation of the general formulas for the screened self-energy correction was presented in
Ref.~\cite{artemyev:05:pra}. We here rearrange the formulas in a form suitable
for a numerical evaluation.
The screened self-energy correction is conveniently represented as a sum of the perturbed-orbital (po),
reducible (red), and vertex (ver) contributions,
\begin{align}\label{eq:se:0}
\Delta E_{\rm sescr} = \Delta E_{\rm sescr, po} + \Delta E_{\rm sescr, red} + \Delta E_{\rm sescr, ver}\,.
\end{align}
The perturbed-orbital contribution is expressed in terms of diagonal and non-diagonal
matrix elements of the one-loop self-energy operator
$ \Sigma(\vare)$,
\begin{widetext}
\begin{align} \label{eq:se:1}
\Delta E_{\rm sescr, po} = &\ F_i\,F_k \sum_{PQ}(-1)^{P+Q}
\bigg[
2 \sum_{n \neq Pi_1}
  \lbr Pi_1 | \Sigma(\vare_{Pi_1})| n\rbr\,
  \frac{\lbr n Pi_2|I(\Delta_{Qk_2,Pi_2})|Qk_1Qk_2\rbr}{\vare_{Pi_1}-\vare_{n}}
\nonumber \\ &
 + \lbr Pi_1 | \Sigma(\vare_{Pi_1})| Pi_1\rbr\,
  \lbr Pi_1 Pi_2|I'(\Delta_{Qk_2,Pi_2})|Qk_1Qk_2\rbr
\bigg] \,,
\end{align}
where the summation over $n$ is performed over the complete Dirac spectrum.
The reducible part of the screened self-energy correction contains the derivative of the self-energy operator and is given by
\begin{align} \label{eq:se:2}
\Delta E_{\rm sescr, red} = &\ F_i\,F_k \sum_{PQ}(-1)^{P+Q}
 \lbr Pi_1 | \Sigma{}'(\vare_{Pi_1})| Pi_1\rbr\,
  \lbr Pi_1 Pi_2|I(\Delta_{Qk_2,Pi_2})|Qk_1Qk_2\rbr
 \,.
\end{align}
The vertex  part of the screened self-energy correction is
\begin{align} \label{eq:se:4}
\Delta E_{\rm sescr, ver} = &\ F_i\,F_k \sum_{PQ}(-1)^{P+Q}
\frac{i}{2\pi}
  \intinf d\omega \,
\sum_{n_1n_2}
\frac{\lbr Pi_1n_2| I(\omega)|n_1Qk_1\rbr\, \lbr n_1Pi_2|I(\Delta_{Qk_2,Pi_2})|n_2Qk_2\rbr}
  {(\vare_{Pi_1}-\omega-u\,\vare_{n_1})(\vare_{Qk_1}-\omega-u\,\vare_{n_2})} \,.
\end{align}
\end{widetext}

The numerical evaluation of the screened self-energy corrections is based on
the use of the analytical representation of the Dirac-Coulomb Green function in terms of
the Whittaker functions, see Ref.~\cite{yerokhin:20:green} for details.
The perturbed-orbital contribution $\Delta E_{\rm po}$ is expressed in
terms of diagonal and non-diagonal matrix elements of the one-loop self-energy operator
$\Sigma(\vare)$, which are computed by numerical methods described in detail in
Refs.~\cite{yerokhin:99:pra,yerokhin:05:se}. The general scheme of evaluation of the reducible
and vertex corrections was developed in Ref.~\cite{yerokhin:99:sescr}. The
contributions of the free electron propagators are separated out and calculated in the
momentum space, after a covariant regularization of ultraviolet divergences and an explicit
cancellation of divergent terms.
The remaining (many-potential) contributions contain infrared-divergent
terms, arising when the energy differences in the energy denominators
vanish (e.g., in Eq.~(\ref{eq:se:4}) with $n_1 = Pi_1$ and $n_2 = Qk_1$).
The divergent terms in the vertex and reducible contributions are
separated out and regularized by introducing a finite photon mass.
The divergencies cancel out in the sum of the vertex and the reducible contributions
(see Appendix B of Ref.~\cite{yerokhin:20:green} for details),
leaving a finite remainder to be calculated numerically.

The many-potential reducible
contribution was computed as a derivative of the one-loop self-energy operator by the
method developed in Ref.~\cite{yerokhin:05:se}. The many-potential vertex contribution contains two bound
electron propagators and represents the main computational difficulty. In the present work, we
evaluate it with the technique described in detail in Ref.~\cite{yerokhin:20:green},
which was recently employed for the evaluation of the self-energy screening corrections
to the $g$ factor \cite{yerokhin:20:gfact}. The expansion over the partial waves in the vertex term
is the main source of the numerical uncertainty. It was extended up to
$|\kappa_{\rm max}| = 50$, with the remainder of the tail estimated by a polynomial
fitting of the expansion terms in $1/|\kappa|$.

Numerical results of our computations of the screened self-energy correction are presented in
Table~\ref{tab:sescr} in terms of the scaled function $G_{\rm
sescr}(\Za)$, with the leading $\alpha$ and $\Za$ dependence pulled out,
\begin{align}\label{eq:se:5}
  \Delta E_{\rm sescr} = m\alpha^2 (\Za)^3\,G_{\rm sescr}(\Za)\,.
\end{align}
Results in Table~\ref{tab:sescr} are obtained for the point nuclear charge model and, unless
explicitly specified, in the Feynman gauge. For $Z = 20$, results are presented also
for the Coulomb gauge for the photon connecting the two electrons.
In the case of the mixing $(1s2p_j)_1$ configurations, contributions to the matrix elements of
the effective Hamiltonian in the $jj$-coupling are presented, see Sec.~\ref{sec:mix}.
The off-diagonal $(1s2p_j)_1$ matrix elements
are listed under the label ``offdiag''.

The comparison presented in Table~\ref{tab:sescr} shows that
our present results are in good agreement with those from previous computations
\cite{yerokhin:97:pla,artemyev:05:pra} but are more accurate. It is also demonstrated that our
numerical results are gauge invariant well within the estimated numerical uncertainty.

\section{Screened vacuum-polarization}
\label{sec:vp}

\begin{table*}[t]
\caption{The screened
vacuum-polarization correction for the $n = 1$ and $n = 2$ states of He-like ions,
in units $\alpha^2(\Za)^3$.
\label{tab:vpscr}}
\begin{ruledtabular}
\begin{tabular}{lw{3.8}w{3.8}w{3.8}w{3.8}w{3.8}w{3.8}w{3.8}w{2.10}l}
\multicolumn{1}{l}{$Z$}
& \multicolumn{1}{c}{$(1s1s)_0$}
  & \multicolumn{1}{c}{$(1s2s)_0$}
    & \multicolumn{1}{c}{$(1s2s)_1$}
        & \multicolumn{1}{c}{$(1s2p_{1/2})_0$}
            & \multicolumn{1}{c}{$(1s2p_{1/2})_1$}
                & \multicolumn{1}{c}{$(1s2p_{3/2})_1$}
                    & \multicolumn{1}{c}{$(1s2p_{3/2})_2$}
                        & \multicolumn{1}{c}{offdiag}
                        & \multicolumn{1}{c}{}
  \\\hline\\[-5pt]
 10  & 0.117\,77         & 0.024\,78         & 0.016\,87         & 0.006\,95         & 0.004\,27   & 0.001\,60         & 0.006\,82       & -0.003\,714 \\
 12  & 0.117\,30         & 0.024\,73         & 0.016\,74         & 0.006\,94         & 0.004\,27   & 0.001\,59         & 0.006\,76       & -0.003\,679 \\
 14  & 0.117\,04         & 0.024\,73         & 0.016\,65         & 0.006\,95         & 0.004\,27   & 0.001\,59         & 0.006\,70       & -0.003\,650 \\
 16  & 0.116\,99         & 0.024\,78         & 0.016\,58         & 0.006\,97         & 0.004\,29   & 0.001\,59         & 0.006\,65       & -0.003\,626 \\
 18  & 0.117\,13         & 0.024\,88         & 0.016\,53         & 0.007\,02         & 0.004\,31   & 0.001\,59         & 0.006\,61       & -0.003\,607\\[3pt]
 20  & 0.117\,47         & 0.025\,03         & 0.016\,51         & 0.007\,08         & 0.004\,35   & 0.001\,60         & 0.006\,58       & -0.003\,592 \\
     & 0.117             & 0.025             & 0.017             & 0.007             & 0.005       & 0.001             & 0.007           & -0.004
     &Ref.~\cite{artemyev:05:pra}\\[3pt]
 24  & 0.118\,7          & 0.025\,48         & 0.016\,53         & 0.007\,26         & 0.004\,45   & 0.001\,61         & 0.006\,53       & -0.003\,575 \\
 28  & 0.120\,7          & 0.026\,13         & 0.016\,64         & 0.007\,52         & 0.004\,60   & 0.001\,64         & 0.006\,50       & -0.003\,574\\[3pt]
 30  & 0.122\,0          & 0.026\,54         & 0.016\,73         & 0.007\,67         & 0.004\,70   & 0.001\,65         & 0.006\,49       & -0.003\,579 \\
     & 0.122             & 0.026\,6          & 0.016\,8          & 0.007\,7          & 0.004\,6    & 0.001\,8          & 0.006\,7        & -0.003\,5
     &Ref.~\cite{artemyev:05:pra}\\[3pt]
 32  & 0.123\,4          & 0.026\,99         & 0.016\,84         & 0.007\,85         & 0.004\,80   & 0.001\,66         & 0.006\,49       & -0.003\,588 \\
 36  & 0.126\,9          & 0.028\,07         & 0.017\,13         & 0.008\,28         & 0.005\,06   & 0.001\,70         & 0.006\,50       & -0.003\,615 \\
 40  & 0.131\,2          & 0.029\,38         & 0.017\,52         & 0.008\,81         & 0.005\,37   & 0.001\,74         & 0.006\,54       & -0.003\,657 \\
\end{tabular}
\end{ruledtabular}
\end{table*}

The derivation of the general formulas for the screened vacuum-polarization correction was presented in
Ref.~\cite{artemyev:05:pra}. We now summarize the final formulas needed for the actual calculation.
The screened vacuum-polarization correction is conveniently represented as a sum of the
perturbed-orbital (po), and the photon-propagator (ph) contributions,
\begin{align}\label{eq:vp:0}
\Delta E_{\rm vpscr} = \Delta E_{\rm vpscr, po} + \Delta E_{\rm vpscr, ph}\,.
\end{align}
\begin{widetext}
The perturbed-orbital contribution is analogous to that for the screened self-energy and is expressed in terms of
matrix elements of the one-loop vacuum-polarization potential,
\begin{align} \label{eq:vp:1}
\Delta E_{\rm vpscr, po} = &\ F_i\,F_k \sum_{PQ}(-1)^{P+Q}
\bigg[
2 \sum_{n \neq Pi_1}
  \lbr Pi_1 | U_{\rm VP}| n\rbr\,
  \frac{\lbr n Pi_2|I(\Delta_{Qk_2,Pi_2})|Qk_1Qk_2\rbr}{\vare_{Pi_1}-\vare_{n}}
\nonumber \\ &
 + \lbr Pi_1 | U_{\rm VP}| Pi_1\rbr\,
  \lbr Pi_1 Pi_2|I'(\Delta_{Qk_2,Pi_2})|Qk_1Qk_2\rbr
\bigg] \,.
\end{align}
The remaining part of the screened vacuum-polarization is given by the correction to the photon
propagator,
\begin{align} \label{eq:vp:2}
\Delta E_{\rm vpscr, ph} = &\ F_i\,F_k \sum_{PQ}(-1)^{P+Q}
  \lbr Pi_1 Pi_2|U_{\rm ph}(\Delta_{Qk_2,Pi_2})|Qk_1Qk_2\rbr
 \,,
\end{align}
where $U_{\rm ph}$ is the radiatively-corrected photon propagator,
\begin{align}
U_{\rm ph}(\delta,\bfx,\bfy) = \frac{\alpha^2}{2\pi i}
 \intinf d\omega \, \int d^3\bfz_1d^3\bfz_2 \,\alpha_{\mu} \,
  D^{\mu\nu}(\delta,\bfx,\bfz_1)\,
 {\rm Tr} \Big[
  \alpha_{\nu}\,G(\omega-\delta/2,\bfz_1,\bfz_2)\,
  \alpha_{\rho}\,G(\omega+\delta/2,\bfz_2,\bfz_1)
  \Big]
  D^{\rho\sigma}(\delta,\bfz_2,\bfy)\,
\alpha_{\sigma}\,
.
\end{align}
\end{widetext}

The screened vacuum-polarization correction was first calculated for the ground state of He-like ions
in Refs.~\cite{persson:96:2el,artemyev:97} and for the $n = 2$ excited states in Ref.~\cite{artemyev:05:pra}. The numerical
calculation of the present work follows the general scheme developed in these studies.
The Wichmann-Kroll part of the vacuum-polarization potential $U_{\rm VP}$ was computed for the point nuclear charge
with help of approximate formulas
obtained in Refs.~\cite{fainshtein:91,manakov:12:vgu,manakov:13:vgu}. The summation over the Dirac spectrum in
Eq.~(\ref{eq:vp:1}) causes no problem and can be computed in different ways. In the present work
we chose to use the $B$-splines basis set method \cite{johnson:88}, which is technically the
simplest choice in this case.

The vacuum-polarization correction to the photon propagator is standardly separated into the Uehling and the
Wichmann-Kroll parts \cite{artemyev:97}. The calculation of the Uehling part is relatively straightforward and
was performed by formulas from Ref.~\cite{artemyev:97}.
The Wichmann-Kroll part of the radiatively-corrected photon propagator
is more difficult to compute but its contribution is very
small for the range of the nuclear charges considered in the present work.
We therefore exclude this correction from our calculation,
estimating its upper bound to make sure that it does not contribute on the level of our present interest.
For the largest $Z$ considered in this work,  $Z = 40$, we estimate the contribution to the function
$G_{\rm vpscr}$ to be less than $1\times 10^{-4}$ for the $(1s)^2$ state and less than
$1\times 10^{-5}$ for the $n = 2$ states \cite{artemyev:20:priv}. As $Z$ decreases, the contribution
diminishes quickly and can be completely ignored in the context of the present investigation.

Our numerical results for the screened vacuum-polarization correction are presented in
Table~\ref{tab:vpscr} in terms of the scaled function $G_{\rm
vpscr}(\Za)$ defined as
\begin{align}\label{eq:vp:5}
  \Delta E_{\rm vpscr} = m\alpha^2 (\Za)^3\,G_{\rm vpscr}(\Za)\,.
\end{align}
The comparison presented in the table shows that our results are in agreement with previous calculations
\cite{artemyev:97,artemyev:05:pra}.

\section{Two-photon exchange}
\label{sec:2ph}

The two-photon exchange correction was first calculated for the ground state of He-like ions in
Ref.~\cite{blundell:93:b}. Calculations for the $n = 2$ excited states were performed independently
by several groups over the course of several decades
\cite{mohr:00:pra,andreev:01,asen:02,andreev:04,artemyev:05:pra,kozhedub:19}.

We here follow the approach of Ref.~\cite{artemyev:05:pra} and express the two-photon exchange correction
as a sum of the irreducible (ir) and the reducible (red) contributions,
\begin{align}\label{eq:2ph:0}
\Delta E_{\rm 2ph} = \Delta E_{\rm 2ph, ir} + \Delta E_{\rm 2ph, red}\,.
\end{align}
\begin{widetext}
The irreducible part is given by
\begin{align} \label{eq:2ph:1}
\Delta E_{\rm 2ph, ir} = &\ F_i\,F_k \sum_{P}(-1)^{P}\frac{i}{2\pi}
 \intinf d\omega\,
    \Bigg\{
    \sum_{n_1n_2}^{E_n\neq E^{(0)}}
    \frac{\lbr Pi_1Pi_2| I(\omega)|n_1n_2\rbr\, \lbr n_1n_2|I(\omega-\Delta_{k_1,Pi_1})|k_1k_2\rbr}
         {(\vare_{Pi_1}+\omega - u\vare_{n_1})(\vare_{Pi_2}-\omega-u\vare_{n_2})}
\nonumber \\ &
   + \sum_{n_1n_2}
    \frac{\lbr Pi_1 n_2| I(\omega)|n_1k_2\rbr\, \lbr n_1Pi_2|I(\omega-\Delta_{k_1,Pi_1})|k_1n_2\rbr}
         {(\vare_{Pi_1}+\omega - u\vare_{n_1})(\vare_{k_2}+\omega-u\vare_{n_2})}
\Bigg\}\,,
\end{align}
where $E_n = \vare_{n_1}+\vare_{n_2}$ and $E^{(0)}$ is the energy of the reference
state. The condition $E_n\neq E^{(0)}$ for excluding intermediate states from the summation over
$n_1$ and $n_2$ in the above formula should be understood differently for the case of non-mixing and the case
of mixing states. In the former case (in this work, for all states except $2\,^1P_1$ and $2\,^3P_1$), the
reference state correspond to an isolated level, with $E^{(0)} =
\vare_{i_1}+\vare_{i_2} = \vare_{k_1}+\vare_{k_2}$. The terms to be excluded from the summation are thus
$(n_1,n_2) = (i_1,i_2)$ and $(i_2,i_1)$. In the latter case, the reference state
belongs to a sub-space of quasidegenerate levels (in this work, the $2\,^1P_1$ and $2\,^3P_1$ states)
and one should exclude all of them from the summation,  $(n_1,n_2) = (1s,2p_{1/2})$,  $(1s,2p_{3/2})$,
$(2p_{1/2},1s)$, $(2p_{3/2},1s)$.

The reducible part accounts for the terms excluded from the summation over $n_1$ and $n_2$ in the irreducible part.
It is given by
\begin{align} \label{eq:2ph:2}
\Delta E_{\rm 2ph, red} = &\ F_i\,F_k \sum_{P}(-1)^{P}\frac{i}{4\pi}
 \intinf d\omega\,
    \sum_{n_1n_2}^{E_n = E^{(0)}}
    \lbr Pi_1Pi_2| I(\omega)|n_1n_2\rbr\, \lbr n_1n_2|I(\omega-\Delta_{k_1,Pi_1})|k_1k_2\rbr
 \nonumber \\ & \times
    \bigg[
    \frac1{(\vare_{Pi_1}+\omega - \vare_{n_1} + i0)(\vare_{Pi_2}-\omega-\vare_{n_2}-i0)}
+    \frac1{(\vare_{Pi_1}+\omega - \vare_{n_1} - i0)(\vare_{Pi_2}-\omega-\vare_{n_2}+i0)}
\bigg]
\,.
\end{align}
\end{widetext}
It might be noted that the terms in the reducible part have the same form as the terms excluded from the
summation in the irreducible part
but differ from them by signs of the infinitesimal imaginary additions in the pole positions.
Note also that formulas for the quasidegenerate states have some arbitrariness in them, which corresponds
to a possibility to re-assign small contributions to higher orders of perturbation theory
\cite{shabaev:02:rep}. In particular, authors of Ref.~\cite{kozhedub:19} used a different choice
of the reference-state energy. This arbitrariness leads to very small numerical changes in the results.

\begin{table*}[t]
\caption{The
two-photon exchange correction for the ground and $n = 2$ states of He-like ions, in eV.
\label{tab:twoph}}
\scriptsize
\begin{ruledtabular}
\begin{tabular}{lw{3.9}w{3.9}w{3.8}w{3.8}w{3.8}w{3.8}w{3.8}w{3.9}l}
\multicolumn{1}{l}{$Z$}
& \multicolumn{1}{c}{$(1s1s)_0$}
  & \multicolumn{1}{c}{$(1s2s)_0$}
    & \multicolumn{1}{c}{$(1s2s)_1$}
        & \multicolumn{1}{c}{$(1s2p_{1/2})_0$}
            & \multicolumn{1}{c}{$(1s2p_{1/2})_1$}
                & \multicolumn{1}{c}{$(1s2p_{3/2})_1$}
                    & \multicolumn{1}{c}{$(1s2p_{3/2})_2$}
                        & \multicolumn{1}{c}{offdiag}
                        & \multicolumn{1}{c}{}
  \\\hline\\[-5pt]
 10& -4.380\,311\,(7)  & -3.156\,509\,(4) & -1.296\,332\,(2) & -2.030\,796\,(3) & -2.769\,646\,(1)& -3.526\,412\,(2) & -1.993\,318\,(2) & -1.073\,247\,(7)\\
 12& -4.419\,431\,(11) & -3.174\,334\,(5) & -1.299\,108\,(2) & -2.050\,558\,(5) & -2.778\,965\,(2)& -3.533\,294\,(2) & -1.996\,368\,(3) & -1.071\,185\,(4)\\
 14& -4.465\,463\,(15) & -3.195\,431\,(7) & -1.302\,403\,(3) & -2.074\,088\,(6) & -2.790\,022\,(2)& -3.541\,435\,(3) & -1.999\,973\,(3) & -1.068\,737\,(5)\\
 16& -4.518\,375\,(18) & -3.219\,828\,(8) & -1.306\,222\,(3) & -2.101\,472\,(7) & -2.802\,838\,(3)& -3.550\,839\,(3) & -2.004\,131\,(4) & -1.065\,905\,(9)\\
 18& -4.578\,150\,(23) & -3.247\,563\,(10)& -1.310\,572\,(4) & -2.132\,807\,(8) & -2.817\,439\,(3)& -3.561\,515\,(4) & -2.008\,838\,(5) & -1.062\,686\,(6)\\[3pt]
 20& -4.644\,780\,(27) & -3.278\,681\,(12)& -1.315\,460\,(5) & -2.168\,207\,(9) & -2.833\,853\,(3)& -3.573\,469\,(4) & -2.014\,093\,(5) & -1.059\,080\,(7) \\
   & -4.644\,769\,(18) & -3.278\,679\,(12)& -1.315\,460\,(5) & -2.168\,207\,(9) & -2.833\,853\,(5)& -3.573\,468\,(9) & -2.014\,093\,(5) & -1.059\,079\,(13)& Coulomb\\
   & -4.644\,76\,(19)  & -3.278\,67\,(3)  & -1.315\,457\,(4) & -2.168\,20\,(2)  & -2.833\,85\,(2) & -3.573\,46\,(5)  & -2.014\,088\,(10)& -1.059\,07\,(5)  & Ref.~\cite{kozhedub:19} \\
   & -4.643\,5         & -3.278\,4        & -1.315\,4        & -2.168\,2        & -2.833\,7       & -3.573\,3        & -2.014\,1        & -1.058\,9      & Ref.~\cite{artemyev:05:pra} \\[3pt]
 24& -4.798\,626\,(37) & -3.351\,293\,(16)& -1.326\,883\,(6) & -2.251\,724\,(10)& -2.872\,262\,(4)& -3.601\,255\,(5) & -2.026\,227\,(6) & -1.050\,710\,(8)\\
 28& -4.980\,080\,(50) & -3.438\,206\,(21)& -1.340\,564\,(8) & -2.353\,222\,(11)& -2.918\,384\,(4)& -3.634\,290\,(6) & -2.040\,495\,(6) & -1.040\,797\,(10)\\[3pt]
 30& -5.081\,262\,(57) & -3.487\,234\,(23)& -1.348\,279\,(9) & -2.411\,161\,(12)& -2.944\,458\,(4)& -3.652\,811\,(7) & -2.048\,415\,(6) & -1.035\,263\,(10)\\
   & -5.081\,22\,(17)  & -3.487\,20\,(5)  & -1.348\,270\,(3) & -2.411\,14\,(2)  & -2.944\,45\,(5) & -3.652\,80\,(10) & -2.048\,399\,(10)& -1.035\,21\,(3)  & Ref.~\cite{kozhedub:19}  \\
   & -5.079\,5         & -3.486\,8        & -1.348\,3        & -2.411\,1        & -2.944\,3       & -3.652\,5        & -2.048\,4        & -1.035\,0      & Ref.~\cite{artemyev:05:pra} \\
   & -5.081\,18        & -3.487\,164      & -1.348\,268      &                  &                 &                  &                  &                   & Refs.~\cite{lindgren:95:pra,asen:02} \\[3pt]
 32& -5.189\,494\,(65) & -3.540\,110\,(26)& -1.356\,593\,(10)& -2.474\,167\,(13)& -2.972\,615\,(4)& -3.672\,691\,(7) & -2.056\,851\,(6) & -1.029\,347\,(11)\\
 36& -5.427\,426\,(82) & -3.657\,867\,(32)& -1.375\,072\,(12)& -2.616\,323\,(14)& -3.035\,439\,(5)& -3.716\,601\,(9) & -2.075\,246\,(7) & -1.016\,373\,(13)\\
 40& -5.694\,644\,(99) & -3.792\,523\,(38)& -1.396\,126\,(14)& -2.781\,799\,(16)& -3.107\,442\,(5)& -3.766\,186\,(10)& -2.095\,623\,(7) & -1.001\,890\,(15)\\
\end{tabular}
\end{ruledtabular}
\end{table*}

The numerical evaluation of the two-photon exchange corrections is rather involved, but relatively well established at
present. Unlike the self-energy correction computed with help of the analytical representation of the Dirac-Coulomb Green
function, all previous computations of the two-photon exchange
\cite{mohr:00:pra,andreev:01,asen:02,andreev:04,artemyev:05:pra,kozhedub:19}
were performed using the spectral representation of the Green function,
with help of the $B$-spline finite basis set method \cite{johnson:88}. The reason behind this is
purely technical. First, the radial integrations are more efficiently computed in the spectral representation.
Second, the computation of the $\omega$ integration along the contour extending near
poles of the electron propagator is easier implemented for the spectral representation of the Green function.

In the present work we follow the previous studies and use  the $B$-spline finite basis set method \cite{johnson:88} for
the evaluation of the two-photon exchange correction. We note that we do not use the
the dual-kinetical-balance (DKB) basis set \cite{shabaev:04:DKB} since we perform calculations for the point
nuclear charge, while the DKB method is formulated for an extended nuclear-charge model only.
The numerical procedure used in this work is very similar to the one
developed for the two-photon exchange correction to the bound-electron $g$ factor and described in detail in
Ref.~\cite{yerokhin:21:gfact}. The computation was performed in two gauges, the Feynman and the Coulomb one. The results
obtained in the two gauges agree well within the estimated numerical uncertainty.

The computation of the two-photon exchange correction needs to be performed to a very high numerical accuracy,
because of the presence of the nonrelativistic ($\sim (\Za)^0$) and relativistic ($\sim (\Za)^2$) contributions.
The QED part of the two-photon exchange enters in the order $(\Za)^3$ only; its identification thus entails severe
numerical cancellations in the low-$Z$ region. By contrast, the screened self-energy and vacuum-polarization corrections
scale as $(\Za)^3$, so their numerical uncertainty is less crucial for the determination of the total two-electron QED
correction.

The dominant sources of numerical uncertainty for the two-photon exchange correction are the convergence with respect
to the number of basis functions $N$ and the truncation of the partial-wave expansion.
Our calculations were performed typically
for several sets of $B$-splines with $N$ up to $N = 125$ and
then extrapolated to $N\to \infty$. The infinite partial-wave summation over the
relativistic angular momentum quantum number $\kappa$ was extended up to $|\kappa_{\rm max} | = 25$, with the remaining
tail estimated by the polynomial fitting of the expansion terms in $1/|\kappa|$.

Numerical results of our calculation are summarized in Table~\ref{tab:twoph}.
They are obtained for the point nuclear charge distribution and, unless
explicitly specified, the Feynman gauge. For $Z = 20$, we present in addition results obtained
in the Coulomb gauge.
The comparison presented in the table shows
that our values are in good agreement with results obtained previously but are more accurate.

\begin{table}
\caption{Coefficients of the $\Za$ expansion of the two-photon exchange correction.
\label{tab:coef:2ph:nonmix}}
\begin{ruledtabular}
\begin{tabular}{lw{2.9}w{2.9}}
\multicolumn{1}{l}{}
& \multicolumn{1}{c}{$a_{20}$}
    & \multicolumn{1}{c}{$a_{40}$}
  \\
\hline\\[-5pt]
$1\,^1 S_0$       & -0.157\,666\,43 & -0.636\,506\,9 \\
$2\,^1 S_0$       & -0.114\,510\,14 & -0.281\,858\,6 \\
$2\,^3 S_1$       & -0.047\,409\,30 & -0.042\,775\,5 \\
$2\,^1 P_1$       & -0.157\,028\,66 & -0.090\,632\,2 \\
$2\,^3 P_0$       & -0.072\,998\,98 & -0.303\,523\,4 \\
$2\,^3 P_1$       & -0.072\,998\,98 & -0.162\,129\,4 \\
$2\,^3 P_2$       & -0.072\,998\,98 & -0.047\,315\,7 \\
$(2\,^1 P_1,2\,^3 P_1)$
      & \multicolumn{1}{c}{0}  & -0.004\,718\,5  \\
\end{tabular}
\end{ruledtabular}
\end{table}

\begin{table*}
\caption{Coefficients of the $\Za$ expansion of the one-loop two-electron QED correction.
\label{tab:coef:nonmix}}
\begin{ruledtabular}
\begin{tabular}{lw{1.7}w{1.6}cw{1.5}w{1.5}w{1.5}}
\multicolumn{1}{l}{}
        & \multicolumn{1}{c}{$a_{51}$}
            & \multicolumn{1}{c}{$a_{50}$}
                & \multicolumn{1}{c}{$a_{61}$}
                & \multicolumn{1}{c}{$a_{60}$}
                    & \multicolumn{1}{c}{$a_{72}$}
                    & \multicolumn{1}{c}{$a_{71}$}
  \\
\hline\\[-5pt]
$1\,^1 S$   &  -0.659\,550\,48  &  1.658\,816\,0 &$\nicefrac{1}{16}$ & -4.3711 &  0.425\,033  &    \\
%
$2\,^1 S$   &  -0.137\,744\,61  &  0.325\,517\,0 &$\nicefrac{1}{81}$ & -1.1041 &  0.089\,554  &         \\
%
$2\,^3 S$   &  -0.089\,756\,44  &  0.191\,147\,5 &   0               & -0.6514 &  0.067\,317  & -0.4829  \\
%
%
$2\,^1 P_1$ & 0.003\,158\,46  & -0.017\,559\,8   &$\nicefrac{1}{243}$&  0.0062 & -0.006\,954 &          \\
$2\,^3 P_0$ &  -0.036\,478\,76  &  0.106\,210\,2 &   0               & -0.6796 &  0.027\,359  & -0.3681  \\
%
%
$2\,^3 P_1$ & -0.036\,478\,76  &  0.081\,192\,9  &   0               & -0.3088 &  0.027\,359  & -0.3171  \\
%
%
$2\,^3 P_2$ &  -0.036\,478\,76  &  0.059\,452\,5 &   0               & -0.2117 &  0.027\,359  & -0.2556  \\
%
%
$(2\,^1 P_1,2\,^3 P_1)$
           &\multicolumn{1}{c}{0} &  0.011\,962\,3 &   0       \\
\end{tabular}
\end{ruledtabular}
\end{table*}

\begin{figure*}
\centerline{
\resizebox{.95\textwidth}{!}{%
  \includegraphics{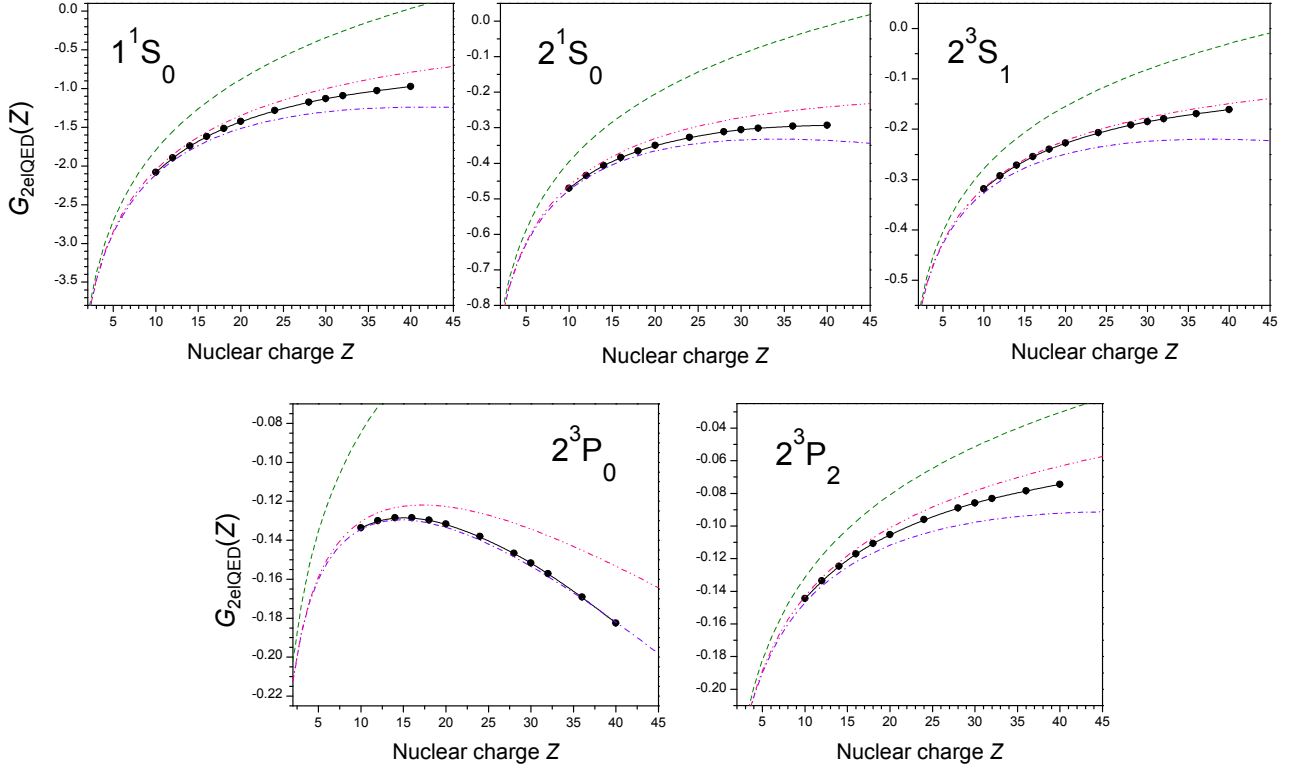}
}}
 \caption{The two-electron QED correction for non-mixing states of He-like ions, as
 a function of the nuclear charge number $Z$, in terms of the function $G_{\rm 2elQED}$
 defined by Eq.~(\ref{eq3}). The dots and solid line (black) present results of
 all-order numerical calculation. The dashed line (green) shows the contribution of the leading
 $\Za$-expansion term
 of order $\alpha^2(\Za)^3$; the dashed-dotted line (violet) is the contribution of two first
 $\Za$-expansion terms of order $\alpha^2(\Za)^3$ and $\alpha^2(\Za)^4$; the double-dotted dashed line
 (pink) shows the contribution of two first $\Za$-expansion terms plus the hydrogenic estimate of
 the higher-order terms.
 \label{fig:1}}
\end{figure*}

\begin{table}
\caption{The higher-order QED and finite nuclear size (FNS) contributions for the ground and the non-mixing
$n = 2$ states. Units are $m\alpha^2(\Za)^5$.
\label{tab:ho:nonmix}}
\begin{ruledtabular}
\begin{tabular}{lw{1.3}w{1.3}w{1.3}w{1.3}w{1.3}w{5.5}}
\multicolumn{1}{l}{$Z$}
& \multicolumn{1}{c}{$1/Z^0$}
    & \multicolumn{1}{c}{$1/Z^1$}
        & \multicolumn{1}{c}{$1/Z^{2+}$}
            & \multicolumn{1}{c}{FNS}
                & \multicolumn{1}{c}{Total}
  \\
\hline\\[-5pt]
\multicolumn{1}{l}{$1^1S$} \\ 
  6 &    -67.917 &      7.105 &     -0.670 &      9.354 &   -52.13\,(60) \\
  8 &    -78.259 &      6.447 &     -0.433 &      9.022 &   -63.22\,(36) \\
 10 &    -87.192 &      5.964 &     -0.308 &      9.411 &   -72.12\,(27) \\
 14 &   -102.405 &      5.304 &     -0.184 &      7.836 &   -89.45\,(17) \\
 18 &   -115.469 &      4.911 &     -0.124 &      7.856 &   -102.83\,(12) \\
 20 &   -121.511 &      4.786 &     -0.106 &      7.533 &   -109.30\,(11) \\
 24 &   -133.013 &      4.641 &     -0.079 &      7.405 &   -121.046\,(91) \\
 28 &   -144.127 &      4.609 &     -0.062 &      7.354 &   -132.226\,(83) \\
 30 &   -149.658 &      4.629 &     -0.056 &      7.731 &   -137.354\,(80) \\[3pt]
\multicolumn{1}{l}{$2^1S$} \\ 
  6 &     -7.967 &      1.320 &     -0.353 &      0.993 &   -6.01\,(33) \\
  8 &     -9.182 &      1.168 &     -0.230 &      1.002 &   -7.24\,(21) \\
 10 &    -10.246 &      1.079 &     -0.164 &      1.075 &   -8.26\,(15) \\
 14 &    -12.119 &      0.989 &     -0.098 &      0.925 &   -10.303\,(87) \\
 18 &    -13.831 &      0.962 &     -0.066 &      0.948 &   -11.987\,(58) \\
 20 &    -14.670 &      0.966 &     -0.056 &      0.917 &   -12.843\,(49) \\
 24 &    -16.380 &      1.001 &     -0.042 &      0.917 &   -14.504\,(38) \\
 28 &    -18.197 &      1.070 &     -0.033 &      0.925 &   -16.236\,(30) \\
 30 &    -19.170 &      1.115 &     -0.030 &      0.980 &   -17.104\,(28) \\[3pt]
\multicolumn{1}{l}{$2^3S$} \\ 
  6 &     -7.967 &      2.041 &     -0.122 &      1.090 &   -4.958\,(41) \\
  8 &     -9.182 &      1.757 &     -0.078 &      1.075 &   -6.428\,(34) \\
 10 &    -10.246 &      1.565 &     -0.055 &      1.137 &   -7.601\,(31) \\
 14 &    -12.119 &      1.321 &     -0.033 &      0.963 &   -9.868\,(18) \\
 18 &    -13.831 &      1.179 &     -0.022 &      0.978 &   -11.696\,(14) \\
 20 &    -14.670 &      1.131 &     -0.019 &      0.943 &   -12.615\,(13) \\
 24 &    -16.380 &      1.064 &     -0.014 &      0.938 &   -14.391\,(12) \\
 28 &    -18.197 &      1.028 &     -0.011 &      0.943 &   -16.237\,(12) \\
 30 &    -19.170 &      1.019 &     -0.010 &      0.998 &   -17.163\,(11) \\[3pt]
\multicolumn{1}{l}{$2^3P_0$} \\ 
  6 &      0.615 &      0.280 &     -0.234 &     -0.128 &   0.53\,(26) \\
  8 &      0.707 &      0.249 &     -0.150 &     -0.094 &   0.71\,(18) \\
 10 &      0.763 &      0.243 &     -0.106 &     -0.078 &   0.82\,(13) \\
 14 &      0.764 &      0.276 &     -0.063 &     -0.045 &   0.931\,(77) \\
 18 &      0.602 &      0.351 &     -0.043 &     -0.034 &   0.876\,(53) \\
 20 &      0.450 &      0.401 &     -0.036 &     -0.028 &   0.787\,(46) \\
 24 &     -0.016 &      0.525 &     -0.027 &     -0.020 &   0.462\,(35) \\
 28 &     -0.732 &      0.679 &     -0.021 &     -0.014 &   -0.088\,(29) \\
 30 &     -1.196 &      0.768 &     -0.019 &     -0.012 &   -0.460\,(26) \\[3pt]
\multicolumn{1}{l}{$2^3P_2$} \\ 
  6 &      0.381 &      0.661 &     -0.234 &     -0.129 &   0.68\,(11) \\
  8 &      0.456 &      0.559 &     -0.151 &     -0.095 &   0.769\,(79) \\
 10 &      0.518 &      0.489 &     -0.107 &     -0.080 &   0.821\,(60) \\
 14 &      0.625 &      0.400 &     -0.063 &     -0.048 &   0.914\,(36) \\
 18 &      0.711 &      0.348 &     -0.043 &     -0.037 &   0.979\,(25) \\
 20 &      0.748 &      0.329 &     -0.036 &     -0.032 &   1.009\,(21) \\
 24 &      0.812 &      0.301 &     -0.027 &     -0.026 &   1.060\,(16) \\
 28 &      0.865 &      0.282 &     -0.021 &     -0.022 &   1.103\,(13) \\
 30 &      0.888 &      0.275 &     -0.019 &     -0.022 &   1.121\,(12) \\
\end{tabular}
\end{ruledtabular}
\end{table}

\section{Higher-order QED: non-mixing states}
\label{sec:nonmix}

We now turn to comparing our present calculations performed to all orders in $Z\alpha$ with results obtained
in the framework of the $\Za$ expansion. After the agreement of the two approaches is demonstrated,
we will proceed to indentifying the higher-order QED remainder
that can be added to the results of the NRQED calculations of Ref.~\cite{yerokhin:10:helike}.
We start with the non-mixing states, i.e., the ground and all $n = 2$ states except $2\,^1P_1$ and $2\,^3P_1$.

\subsection{Comparison with $\bm \Za$ expansion}

In order to identify the QED part of our all-order results, we
need first to remove the nonrelativistic and relativistic contributions
from the two-photon exchange correction. The QED part of the two-photon exchange
is given by the function $G_{\rm 2ph}$ defined as
\begin{align}\label{eq1}
  \Delta E_{\rm 2ph} = m\alpha^2 \left[ a_{20} + (\Za)^2\,a_{40} + (\Za)^3\,G_{\rm 2ph}(\Za)\right]\,.
\end{align}
Here the coefficient $a_{20}$ arises from the $Z^{-2}$ term of the $1/Z$ expansion of the nonrelativistic
energy, whereas $a_{40}$ comes from the $Z^{-2}$ term of the $1/Z$ expansion of the Breit
correction $\sim\!m\alpha^4$. The function $G_{\rm 2ph}(\Za)$ contains terms of order
$m\alpha^5$ and higher.
Numerical results for the coefficients $a_{20}$ and $a_{40}$ were obtained in Ref.~\cite{yerokhin:10:helike}
and are summarized in Table~\ref{tab:coef:2ph:nonmix}.

The total two-electron QED correction is defined as the sum of the self-energy, vacuum-polarization, and two-photon-exchange
parts,
\begin{align}\label{eq3}
\Delta E_{\rm 2elQED} \equiv &\  m\alpha^2\,(\Za)^3\,G_{\rm 2elQED}(\Za)
 \nonumber \\ = &\
m\alpha^2\,(\Za)^3\,\Big[
G_{\rm sescr} + G_{\rm vpscr} + G_{\rm 2ph} \Big]\,.
\end{align}
It should be noted that the above definition of the two-electron QED contribution includes only
one-loop effects. It does not include the
two-loop correction, for which no all-order calculations have been performed so far.
The $\Za$-expansion of the function $G_{\rm 2elQED}$ reads
\begin{align}\label{eq4}
 G_{\rm 2elQED}(\Za) = &\
   L\,a_{51} + a_{50} + \alpha L\,a_{61}
 \nonumber \\
 &   + (\Za)\, a_{60} + (\Za)^2\, G_{\rm 2elQED}^{(7+)}\,,
\end{align}
where $L \equiv \ln(\Za)^{-2}$ and $G_{\rm 2elQED}^{(7+)}$ is the higher-order remainder,
\begin{align}\label{eq4b}
 G_{\rm 2elQED}^{(7+)}(\Za) =  L^2\,a_{72}
   + L\,a_{71} + a_{70}
 + (\Za) L\,a_{81}+ \ldots \,.
\end{align}
The coefficients $a_{51}$, $a_{50}$, and $a_{61}$
were obtained in Ref.~\cite{yerokhin:10:helike} by fitting the $1/Z$ expansion of the NRQED results.
The coefficient $a_{60}$ was also obtained in Ref.~\cite{yerokhin:10:helike}, but for our present
purposes it needed to be reevaluated to exclude the two-loop contribution.
The coefficient
$a_{72}$ is proportional to the Dirac $\delta$ function and is immediately
obtained from the hydrogen theory. Specifically, $a_{72}$ is induced by the
self-energy coefficient $A_{62}(ns) = -1$, see Eq.~(8) of Ref.~\cite{yerokhin:18:hydr},
\begin{align}\label{eq6}
a_{72} &\ = A_{62}\,\frac{c_1}{\pi} = -\frac{c_1}{\pi}\,,
\end{align}
where $c_1$ is the $1/Z^1$ coefficient of the $1/Z$ expansion of the matrix element of the Dirac
$\delta$ function,
\begin{align}\label{eq7}
\Big< \frac{\pi}{Z^3}\,\big[\delta^3(r_1)+\delta^3(r_2)\big]\Big> = c_0 + \frac{c_1}{Z} + \frac{c_2}{Z^2} + \ldots\,.
\end{align}
The coefficients $c_i$ for the $n = 1$ and $n = 2$ states of He-like ions are given in Table~I of
Ref.~\cite{drake:88:cjp}.
The coefficient $a_{71}$ for the triplet states is evaluated in the present work, on the basis of
formulas derived in Ref.~\cite{patkos:21:helamb}. For the singlet states, $a_{71}$ is unknown.
The nonlogarithmic $m\alpha^7$ correction was evaluated in Ref.~\cite{patkos:21:helamb} for helium;
no results for its $1/Z$ expansion coefficients were reported yet.
The logarithmic coefficient in the next order, $a_{81}$, is also proportional to the Dirac $\delta$ function
and thus can be obtained from the hydrogen theory. Specifically, $a_{81} = (427/192 - \ln 2)\, c_1$
\cite{karshenboim:97,mohr:75:prl}.
The summary of all known $\Za$-expansion coefficients is given in Table~\ref{tab:coef:nonmix}.

In Fig.~\ref{fig:1} we present a comparison of our all-order calculations of the
two-electron QED contribution with predictions based on the $\Za$ expansion. We conclude that the all-order
results in the low-$Z$ region converge to the predictions of the $\Za$ expansion
and that the difference is consistent with the expected magnitude of higher-order effects.

After checking the consistency of our all-order calculations with the known terms of the $\Za$ expansion,
we are now in a position to
obtain numerical results for the higher-order remainder
function $G_{\rm 2elQED}^{(7+)}$ defined by Eq.~(\ref{eq4}), by
subtracting contributions of lower orders in $\Za$ from the all-order results.
Since our numerical calculations are performed for $Z \ge 10$, we have to use an extrapolation
in order to get results for smaller values of $Z$. The extrapolation is complicated by the presence
of logarithms in the expansion of the function. For the triplet states,
we make use of the NRQED results for the logarithmic coefficients $a_{72}$, $a_{71}$, and $a_{81}$
in order to improve the accuracy of the extrapolation. Specifically, we subtract all
known logarithmic terms, apply a polynomial extrapolation, and then re-add the logarithmic
terms back. For the singlet states, the logarithmic coefficient $a_{71}$ is not known,
so we had to include the single logarithmic term into the fitting function. For this reason,
the accuracy of the fit was lower for the singlet than for the triplet states.

\subsection{Higher-order QED contribution}

The calculation of all QED effects
up to order $m\alpha^6$  to energies of the $n = 1$ and
$n = 2$ states of light He-like ions was performed in
Ref.~\cite{yerokhin:10:helike}. These results needed to be complemented
with a separate treatment of the higher-order effects
of order $m\alpha^7$ and higher.

The higher-order QED contribution to the ionization energy of a $1snl_j$ state
was evaluated in Ref.~\cite{yerokhin:10:helike} as
a sum of three parts,
\begin{align}\label{eq8}
E^{(7+)} = E_D^{(7+)} + E_{\rm 1ph}^{(7+)} + E_{\rm rad}^{(7+)}\,,
\end{align}
where the first term is the higher-order
part of the Dirac energy of the $nl_j$ one-electron state,
the second term is the higher-order
part of the one-photon exchange correction,
and the third term is the higher-order part of the radiative QED correction.
The first two terms
are readily evaluated numerically, whereas the radiative QED
contribution was evaluated in Ref.~\cite{yerokhin:10:helike} by rescaling the
hydrogenic result with the expectation value of the Dirac $\delta $ function.
Specifically, the following expression was used:
\begin{align}\label{eq9}
E_{\rm rad}^{(7+)} = &\ \Big[ E_{\rm rad, H}^{(7+)}(1s)+ E_{\rm rad, H}^{(7+)}(nl_j)\Big]
  \nonumber \\ & \times
  \frac{\Big<  \frac{\pi}{Z^3} \big[\delta^3(r_1)+ \delta^3(r_2)\big]\Big>}{1 + \frac{\delta_{l,0}}{n^3}}
  - E_{\rm rad, H}^{(7+)}(1s)\,,
\end{align}
where $E_{\rm rad, H}^{(7+)}(nl_j)$ is the hydrogenic radiative correction of
order of $m\alpha^7$ and higher. The approximation of Eq.~(\ref{eq9}) is
sometimes referred to as the hydrogenic approximation.

Formula (\ref{eq9}) is exact to the leading (zeroth) order in $1/Z$ but only approximate to higher
orders in $1/Z$. In the present work, we calculated the two-electron QED correction
to all orders in $\Za$. With this calculation, we can improve on Eq.~(\ref{eq9}) and make it exact
to the first order in $1/Z$ for the dominant one-loop effects.
We obtain the additional
contribution to be added to Eq.~(\ref{eq9}) as
\begin{align}\label{eq10}
E_{\rm add}^{(7+)} &\ = G_{\rm 2elQED}^{(7+)}
 \nonumber \\ &
-
\Big[ E_{\rm 1loop, H}^{(7+)}(1s)+ E_{\rm 1loop, H}^{(7+)}(nl_j)\Big]
\, \frac{c_1}{Z\Big(1 + \frac{\delta_{l,0}}{n^3}\Big)} \,,
\end{align}
where $G_{\rm 2elQED}^{(7+)}$ is defined by Eq.~(\ref{eq4b}),
$E_{\rm 1loop, H}^{(7+)}$ is the one-loop part of $E_{\rm rad, H}^{(7+)}$, and
$c_1$ is the $1/Z^1$ coefficient of the expansion of the matrix element of the $\delta$ function,
see Eq.~(\ref{eq7}). The subtraction in Eq.~(\ref{eq10}) is needed in order to remove the double
counting of terms already included into the approximation of Eq.~(\ref{eq9}).

In Table~\ref{tab:ho:nonmix} we collect our final results for the higher-order QED correction.
For transparency, we separate the correction into three parts according to the order in $1/Z$ (the
zeroth-, first-, and higher-order contributions). The $1/Z^0$ part is the one-electron
contribution and is given by the sum $E_D^{(7+)}(nl_j) + E_{\rm rad, H}^{(7+)}(nl_j)$,
as tabulated in Refs.~\cite{yerokhin:15:Hlike,yerokhin:18:hydr}.
The $1/Z^1$ term consists of the one-photon exchange correction
$E_{\rm 1ph}^{(7+)}$, the one-loop two-electron QED
contribution $G_{\rm 2elQED}^{(7+)}$, and the two-loop two-electron QED contribution
evaluated within the hydrogenic approximation.
The $1/Z^{2+}$ term is given by the $1/Z^{2+}$ part of Eq.~(\ref{eq9}).

An important point is the estimation of the uncertainty of the higher-order QED correction, because it directly translates
into the uncertainty of the total theoretical energies. The dominant
error comes from the $1/Z^2$ one-loop QED effects. We estimate it
by rescaling the corresponding $1/Z$ correction, with a conservative factor of 1.5,
$$
1.5\, E_{\rm add}^{(7+)}\, \frac{c_2}{Zc_1}\,,
$$
where $c_1$ and $c_2$ are the $\delta$-function expansion coefficients from Eq.~(\ref{eq7}).
Another uncertainty comes from the two-electon two-loop QED effects. It is important that
to the order $m\alpha^7$, the hydrogenic approximation is exact
for the two-loop effects  \cite{patkos:21:helamb}, so that the
uncertainty comes from the $m\alpha^{8+}$ contributions only. We take it to be
100\% of the $1/Z$ $m\alpha^{8+}$ two-loop QED correction, as delivered by the hydrogenic approximation.
Furthermore, we include
the uncertainty due to our extrapolation of $E_{\rm add}^{(7+)}$ (for $Z<10$) and
the uncertainty from the one-electron
two-loop QED corrections \cite{yerokhin:15:Hlike,yerokhin:18:hydr}. Adding all four uncertainties
quadratically, we arrive at the error estimates listed in Table~\ref{tab:ho:nonmix}.

Table~\ref{tab:ho:nonmix} presents also results for the finite nuclear size (fns) correction.
Although its nominal order is $m\alpha^4$, this correction is additionally suppressed by a square of the nuclear
charge radius, which makes it comparable in magnitude to the $m\alpha^{7+}$ QED effects.
Data presented in Table~\ref{tab:ho:nonmix} show that our theoretical predictions are sensitive
to nuclear effects, specifically, to the fns correction. Defining the sensitivity as the ratio of the
theoretical uncertainty in Table~\ref{tab:ho:nonmix} to the fns correction, we find that
the best sensitivity is achieved for the $2\,^3 S$ ionization energy, where it varies from
4\% for $Z = 6$ till 1\% for $Z = 30$.

\section{Higher-order QED: Mixing states}
\label{sec:mix}

Among the $n = 2$ states of He-like ions there are two that have
the same values of the total angular momentum and parity, namely,
the $2\,^1 P_1$ and $2\,^3 P_1$ states. These states strongly mix with each other, especially
for medium-$Z$ ions, and thus should be treated as quasidegenerate.

The QED theory of quasidegenerate states was developed in the framework of the two-time Green function approach
in Refs.~\cite{shabaev:02:rep,artemyev:05:pra}. Within this method, the energies of the $2\,^1 P_1$ and $2\,^3 P_1$
states are determined as eigenvalues of the effective 2$\times$2 Hamiltonian in the $jj$ coupling,
\begin{widetext}
\begin{align}\label{eq:Hjj}
H^{jj} = \Dmatrix{\lbr (1s2p_{\nicefrac{1}{2}})_1| H | (1s2p_{\nicefrac{1}{2}})_1 \rbr}
    {\lbr (1s2p_{\nicefrac{1}{2}})_1| H | (1s2p_{\nicefrac{3}{2}})_1 \rbr}
    {\lbr (1s2p_{\nicefrac{3}{2}})_1| H | (1s2p_{\nicefrac{1}{2}})_1 \rbr}
    {\lbr (1s2p_{\nicefrac{3}{2}})_1| H | (1s2p_{\nicefrac{3}{2}})_1 \rbr}
    \equiv
    \Dmatrix{H_{\nicefrac{1}{2}}}{H_{\nicefrac{1}{2},\nicefrac{3}{2}}}{H_{\nicefrac{3}{2},\nicefrac{1}{2}}}{H_{\nicefrac{3}{2}}}
    \,,
\end{align}
\end{widetext}
with $H_{\nicefrac{3}{2},\nicefrac{1}{2}} = H_{\nicefrac{1}{2},\nicefrac{3}{2}}$.
The all-order calculations described in Secs.~\ref{sec:se}-\ref{sec:2ph} provide results
for the matrix elements of the Hamiltonian $H^{jj}$.

For light ions, the $LS$-coupling scheme yield better results and the $\Za$-expansion calculations are
usually performed within the $LS$-coupling.
For comparing with the $\Za$-expansion calculations, it is convenient to transform the effective
Hamiltonian $H^{jj}$ delivered by the all-order calculations
to the $LS$-coupling by \cite{drake:82:nimb}
\begin{align}\label{eq:HLS}
H^{LS} ={\cal R}\, H^{jj}\, {\cal R}^{-1} \equiv
\Dmatrix{H_T}{H_{ST}}{H_{ST}}{H_S}\,,
\end{align}
where the unitary matrix ${\cal R}$ is
\begin{align}  \label{Rmatrix}
{\cal R}  = \Dmatrix{a}{-b}{b}{a}\,,
\end{align}
with $a = \sqrt{2/3}$ and $b = \sqrt{1/3}$.
The indices $S$ and $T$ in the matrix elements
stand for the singlet ($2\,^1 P$) and triplet ($2\,^3 P$) states, respectively.
The explicit form of the matrix elements is
\begin{align}
\label{eq:mix:02}
H_T =&\
b^2\,H_{\nicefrac{3}{2}}+ a^2\,H_{\nicefrac{1}{2}} - 2ab\,H_{\nicefrac{1}{2},\nicefrac{3}{2}}\,,
\\
\label{eq:mix:03}
H_S =&\  a^2\,H_{\nicefrac{3}{2}} + b^2\,H_{\nicefrac{1}{2}} +
2ab\,H_{\nicefrac{1}{2},\nicefrac{3}{2}}\,,
\\
H_{ST} =&\
ab\,\big[H_{\nicefrac{1}{2}}-H_{\nicefrac{3}{2}}\big] +\big(a^2 - b^2\big)H_{\nicefrac{1}{2},\nicefrac{3}{2}}\,.
\label{eq:mix:01}
\end{align}
We note that the eigenvalues of $H^{jj}$
and $H^{LS}$ are the same since ${\cal R}$ is unitary.

The eigenvalues of $H^{LS}$ denoted as $E_S$ and $E_T$ are obtained as
\begin{align}\label{eq:mix:1}
E_T = H_T + E_{\rm mix}\,,\\
E_S = H_S - E_{\rm mix}\,,
\end{align}
where $E_{\rm mix}$ is the mixing correction
\begin{align}\label{eq:mix:2}
E_{\rm mix} =  \frac{H_T-H_S}{2}\left[ \sqrt{1+\left(\frac{2H_{ST}}{H_T-H_S}\right)^2}-1\right]\,.
\end{align}

For small $Z$, the nondiagonal matrix element $H_{ST}$ is suppressed by a factor of $\alpha^2$ as compared
to the diagonal ones. For this reason $E_{\rm mix}$ is a small correction when $Z \to 0$.
For $|H_{ST}| \ll |H_S-H_T|$ we can expand the square root
in Eq.~(\ref{eq:mix:2}), obtaining
\begin{align}\label{eq:mix:3}
E_{\rm mix} =  \frac{H_{ST}^2}{H_T-H_S} + \frac{H_{ST}^4}{(H_T-H_S)^3} + \ldots.
\end{align}
The terms in the right-hand-side of this formula can be identified as the second-order,
fourth-order, etc perturbation corrections for the single-level perturbation theory. This
shows the equivalence of the perturbation theory for quasidegenerate states and that for
a single level: the perturbation expansion for quasidegenerate states corresponds to a
resummation of the expansion for the single level, accounting for terms with small
energy denominators to all orders.

In Ref.~\cite{yerokhin:10:helike}, the mixing contribution was accounted for within
the lowest order of perturbation theory (as a part of the $m\alpha^6$ correction),
\begin{align}\label{eq:mix:4}
E_{\rm mix}^{(6)} =  \frac{\big[H_{ST}^{(4)}\big]^2}{H^{(2)}_T-H^{(2)}_S} \,,
\end{align}
where $H^{(2)}_T  = E_0(2{\,}^3 P)$ and $H^{(2)}_S  = E_0(2{\,}^1 P)$ are the nonrelativistic energies
of the $2{\,}^3 P$ and $2{\,}^1 P$ states, respectively, and
\begin{align}\label{eq:mix:5}
H_{ST}^{(4)} = \lbr 2{\,}^3 P| H^{(4)} | 2{\,}^1 P \rbr\,,
\end{align}
with $H^{(4)}$ being the Breit Hamiltonian.
In the present work, we extend this formula for $E_{\rm mix}$ by
including higher-order corrections. Specifically, we take $H_{S}$ and $H_{T}$
to be the complete energies of the $2{\,}^1 P$ and $2{\,}^3 P$ states as evaluated in
Ref.~\cite{yerokhin:10:helike} (but without the mixing $m\alpha^6$ contribution) and the nondiagonal
matrix element determined as
\begin{align}\label{eq:mix:6}
H_{ST} = &\ H_{ST}^{(4)} + \alpha\, H_{ST}^{(5)} + \alpha^2\,  H_{ST}^{(6+)}\,.
\end{align}
The $m\alpha^5$ correction to the
nondiagonal matrix element is
\begin{align}\label{eq:mix:7}
H_{ST}^{(5)} = \lbr 2{\,}^3 P| H^{(5)} | 2{\,}^1 P \rbr\,,
\end{align}
where $H^{(5)}$ is the anomalous magnetic moment correction to the Breit Hamiltonian,
given by Eq.~(14) of Ref.~\cite{yerokhin:10:helike}.
Furthermore, $H_{ST}^{(6+)}$ is the correction to the
nondiagonal matrix element of order $m\alpha^6$ and higher. In it 
we retain only the contribution of zeroth order in $1/Z$, which is, according to
Eq.~(\ref{eq:mix:01}),
\begin{align}\label{eq:mix:8}
H_{ST}^{(6+)} = ab\,\Big[E^{(6+)}_{\rm rad,H}(2p_{1/2}) - E^{(6+)}_{\rm rad,H}(2p_{3/2})
 \Big]\,,
\end{align}
where $E^{(6+)}_{\rm rad,H}(nl_j)$ is the one-electron radiative correction
of order $m\alpha^6$ and higher.

\begin{figure}
\centerline{
\resizebox{\columnwidth}{!}{%
  \includegraphics{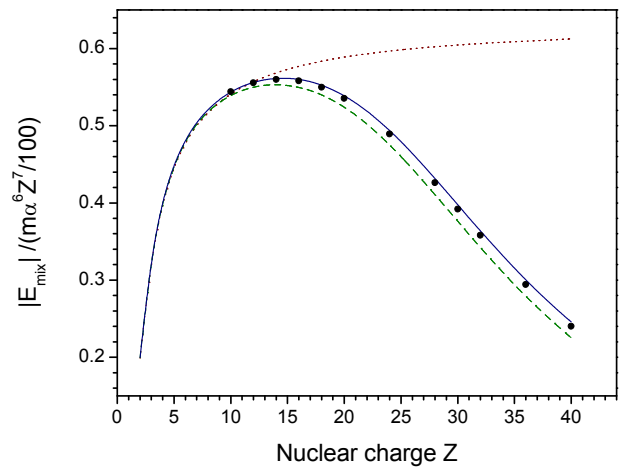}
}}
 \caption{
 The absolute value of the mixing correction $E_{\rm mix}$ divided by $m\alpha^6 Z^7 \times 10^{-2}$. Black dots show
 results of all-order QED calculations, the dotted line (brown) is the contribution of the leading term
 of the $\alpha$ expansion $E_{\rm mix}^{(6)}$; the dashed line (green)
 corresponds to the complete formula (\ref{eq:mix:2}) that includes the next-to-leading terms of the $\alpha$ expansion;
 the solid line (blue) is obtained with the complete formula including all known contributions, see text.
 \label{fig:mix}}
\end{figure}

Fig.~\ref{fig:mix} shows a comparison of the mixing correction $E_{\rm mix}$ evaluated within
the $\Za$-expansion approach in the $LS$ coupling and the corresponding correction obtained with the
all-order approach in the $jj$ coupling (after the coupling transformation according to
Eq.~(\ref{eq:HLS})). We find good agreement of the results obtained with different
approaches. It is clearly seen that the leading-order formula
(\ref{eq:mix:4}) is adequate for very low $Z$ but starts to deviate significantly
from the complete results for $Z > 10$.

A detailed comparison between the all-order and $\Za$-expansion calculations for the mixing states
is complicated by the fact that the matrix elements of the effective Hamiltonian in different methods
are not directly comparable to each other. It is known \cite{malyshev:21:belike}  that the effective Hamiltonian
is defined up to a unitary transformation and the equivalence of different methods is achieved for the eigenvalues
but not for individual matrix elements. More exactly, the equivalence of the matrix elements
exists in the nonrelativistic limit but breaks down on the level of relativistic corrections.
Indeed, we find perfect agreement for the leading coefficient of the $\Za$ expansion
of the two-photon exchange correction $a_{20}$ as obtained from the NRQED calculation
(see Table~\ref{tab:coef:2ph:nonmix}) and from the all-order calculations (after transformation
to the $LS$ coupling with Eq.~(\ref{eq:HLS})). Already for the next-order coefficient $a_{40}$,
however, there is no direct equivalence on the level of individual matrix elements.

While a direct comparison of matrix elements in different methods does not seem to be possible,
one can still \cite{malyshev:21:belike} compare the {\em trace} of the Hamiltonian matrix since it is
preserved by a unitary transformation. Indeed, we find that the sum of the $a_{40}$ coefficients
obtained from the all-order calculation for the $(1s2p_{1/2})_1$ and $(1s2p_{3/2})_1$ diagonal
matrix elements yields $-0.252\,72$, which is close to $-0.252\,762$ obtained for
the sum of the $a_{40}$  coefficients for the $2\,^1P_1$ and  $2\,^3P_1$ diagonal matrix elements
in Table~\ref{tab:coef:2ph:nonmix}.

\begin{figure}
\centerline{
\resizebox{\columnwidth}{!}{%
  \includegraphics{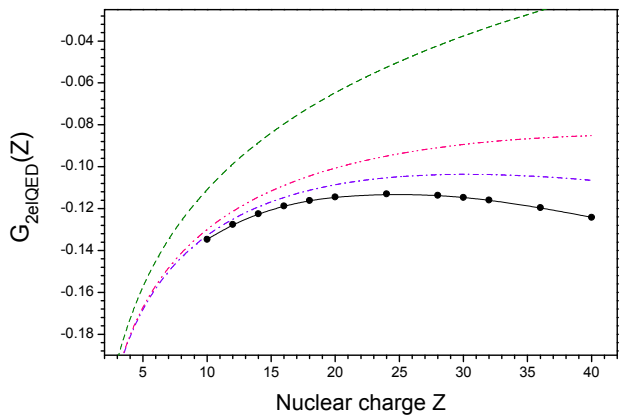}
}}
 \caption{The two-electron QED contribution for the sum of the diagonal
 $(1s2p_{1/2})_1$ and $(1s2p_{3/2})_1$ matrix elements, expressed in terms of the function  $G_{\rm 2elQED}(Z)$
  defined by Eq.~(\ref{eq3}). 
  The line description is the same as for Fig.~\ref{fig:1}.
 \label{fig:mix:trace}}
\end{figure}

In Fig.~\ref{fig:mix:trace} we present a comparison of the two-electron QED contribution for the
sum of the diagonal $(1s2p_{1/2})_1$ and $(1s2p_{3/2})_1$ matrix elements evaluated with
the all-order approach and within the $\Za$-expansion method. The $\Za$-expansion coefficients are taken from
Tables~\ref{tab:coef:2ph:nonmix} and \ref{tab:coef:nonmix}. We observe that the all-order
results converge to predictions of the $\Za$ expansion in the low-$Z$ region
and that the difference is consistent with the expected magnitude of higher-order effects.

Despite good agreement between the two methods, we
presently do not see a way to separate out the higher-order two-electron QED contribution
that can be unambiguously added to $\Za$-expansion results
for the mixing states. The reason is that $E_{\rm mix}$ mixes different
orders of $\Za$ and $1/Z$ expansions so that some double-counting seems to be unavoidable.
For this reason we restrict ourselves to retaining only the one-electron part
of the higher-order QED effects.
Our final results for the higher-order QED correction for the mixing states
are summarized in Table~\ref{tab:ho:mix}. The $1/Z^0$ term is
the sum of the Dirac energy and the radiative contribution.
The radiative part is, according to Eqs.~(\ref{eq:mix:02}) and (\ref{eq:mix:03}),
\begin{align}\label{eq:mix:9}
E_{\rm rad, H}^{(7+)}(2\,^3 P_1) = \frac13\,E_{\rm rad, H}^{(7+)}(2p_{3/2}) + \frac23\,E_{\rm rad, H}^{(7+)}(2p_{1/2})\,,\\
E_{\rm rad, H}^{(7+)}(2\,^1 P_1) = \frac23\,E_{\rm rad, H}^{(7+)}(2p_{3/2}) + \frac13\,E_{\rm rad, H}^{(7+)}(2p_{1/2})\,,
\end{align}
and the same for the Dirac energy $E_D^{(7+)}$. The $1/Z^{1+}$ term contains
the radiative correction within the hydrogenic approximation (\ref{eq9}).
The fourth column shows results for the higher-order mixing correction
$\delta E = E_{\rm mix} - E_{\rm mix}^{(6)}$. The finite nuclear size correction is listed in the fifth column
of the table.

The dominant theoretical uncertainty is caused by the mixing correction. It comes through $H_{ST}^{(6+)}$ and is induced
by contributions of order $1/Z^1$ and higher omitted in Eq.~(\ref{eq:mix:8}).
We estimate the magnitude of the omitted effects
as $(8/Z)\, H_{ST}^{(6+)}$, where the prefactor of $8 = n^3$
accounts for the fact that the $1/Z^1$ contribution is enhanced by the admixture of the $1s$
electron state.

The results collected in Table~\ref{tab:ho:mix} show
that the higher-order mixing correction grows fast with increase of $Z$
and becomes dominant already at $Z = 14$. Correspondingly, the uncertainty due to
missing $1/Z^{1+}$ contributions in the non-diagonal matrix element $H_{ST}^{(6+)}$ becomes
overwhelming for $Z>14$. This reflects the failure of the $LS$ coupling scheme and
the advantage of using the $jj$ coupling for medium and high-$Z$ ions.

\begin{table}
\caption{The higher-order QED and finite nuclear size contributions for the mixing
$n = 2$ states. Units are $m\alpha^2(\Za)^5$.
\label{tab:ho:mix}}
\begin{ruledtabular}
\begin{tabular}{lw{1.3}w{1.3}w{1.3}w{1.3}w{1.3}w{5.5}}
\multicolumn{1}{l}{$Z$}
& \multicolumn{1}{c}{$1/Z^0$}
        & \multicolumn{1}{c}{$1/Z^{1+}$}
        & \multicolumn{1}{c}{MIX}
            & \multicolumn{1}{c}{FNS}
                & \multicolumn{1}{c}{Total}
  \\
\hline\\[-5pt]
\multicolumn{1}{l}{$2\,^1 P_1$} \\ 
  6 &      0.459 &     -0.149 &      0.174 &      0.027 &         0.51\,(21) \\
  8 &      0.539 &     -0.143 &      0.315 &      0.021 &         0.73\,(30) \\
 10 &      0.600 &     -0.131 &      0.311 &      0.019 &         0.80\,(50) \\
 14 &      0.671 &     -0.103 &     -1.930 &      0.012 &        -1.3\,(13) \\
 18 &      0.675 &     -0.071 &    -13.743 &      0.010 &       -13.1\,(25) \\
 20 &      0.649 &     -0.054 &    -27.391 &      0.009 &       -26.8\,(33) \\
 24 &      0.536 &     -0.017 &    -81.309 &      0.008 &       -80.8\,(50) \\
 28 &      0.333 &      0.023 &   -183.583 &      0.008 &      -183.2\,(65) \\
 30 &      0.193 &      0.045 &   -255.344 &      0.009 &      -255.1\,(71) \\[3pt]
\multicolumn{1}{l}{$2\,^3 P_1$} \\ 
  6 &      0.537 &      0.742 &     -0.174 &     -0.129 &         0.98\,(71) \\
  8 &      0.623 &      0.703 &     -0.315 &     -0.094 &         0.92\,(66) \\
 10 &      0.681 &      0.667 &     -0.311 &     -0.079 &         0.96\,(73) \\
 14 &      0.718 &      0.619 &      1.930 &     -0.046 &         3.2\,(14) \\
 18 &      0.638 &      0.602 &     13.743 &     -0.035 &        14.9\,(25) \\
 20 &      0.549 &      0.603 &     27.391 &     -0.029 &        28.5\,(33) \\
 24 &      0.260 &      0.621 &     81.309 &     -0.022 &        82.2\,(50) \\
 28 &     -0.199 &      0.660 &    183.583 &     -0.017 &       184.0\,(65) \\
 30 &     -0.502 &      0.686 &    255.344 &     -0.016 &       255.5\,(71) \\
\end{tabular}
\end{ruledtabular}
\end{table}

\section{Results and discussion}
\label{sec:results}

In Table~\ref{tab:IE:nonmix} we collect our final results for theoretical ionization energies of the ground and the non-mixing
$n = 2$ excited states of helium-like ions with the nuclear charges $Z = 5\,$--$\,30$.
We do not present results for $Z < 5$
since in this region the $1/Z$ expansion employed in the all-order approach ceases to be useful.
For each element, calculations are performed for one isotope with the mass number $A$ specified in the table.
The nuclear masses are obtained from the atomic masses tabulated in Ref.~\cite{wang:12} and the nuclear radii are
taken from Ref.~\cite{angeli:13}. Our results include all non-recoil QED effects up to order $m\alpha^6$ and
the recoil effects up to order $m^2\alpha^5/M$, as calculated for $Z \le 12$ in Ref.~\cite{yerokhin:10:helike}.
For $Z>12$, we sum up the $1/Z$-expansion coefficients listed in Ref.~\cite{yerokhin:10:helike}.
In addition to the NRQED results, we include the higher-order $m\alpha^{7+}$ correction calculated in
this work and summarized in
Table~\ref{tab:ho:nonmix}.

In Table~\ref{tab:IE:nonmix} our present results are
compared with those from our previous work \cite{yerokhin:10:helike} and by other
authors \cite{artemyev:05:pra,malyshev:19,kozhedub:19}. For $Z \geq 12$, our calculation
is in excellent agreement with previous calculations performed within
the all-order approach \cite{artemyev:05:pra,malyshev:19,kozhedub:19} and
improves their accuracy by more than an order of magnitude. For $Z \le 12$, we significantly improve
upon our previous results \cite{yerokhin:10:helike} but also find small deviations in
some cases. The reason for the deviations is that the hydrogenic
approximation used in Ref.~\cite{yerokhin:10:helike}  for estimations of the
higher-order two-electron QED effects turned out to be less accurate than expected. Indeed, it can be seen
from Figs.~\ref{fig:1} and \ref{fig:mix} that the addition of the higher-order correction calculated within
the hydrogenic approximation leads to significant improvements
only in the case of the $2\,^3 S$ state.
For other states, the hydrogenic approximation largely overestimates the actual contribution and even
worsens the results for the $2\,^1P_0$ and $2\,P_1$ states. We conclude that in the case under consideration the
hydrogenic approximation yields only the order of magnitude of the effect but does not provide a
quantitative prediction.

In Table~\ref{tab:IE:mix} we present theoretical energies for the mixing states, $2\,^1 P_1$ and $2\,^3 P_1$. As compared to
our previous investigation \cite{yerokhin:10:helike}, we added the higher-order mixing
correction and re-evaluated the uncertainty. The values of theoretical energies did not change much, but the
uncertainty was increased typically by a factor of 2 or 3. Our results for $Z = 12$ are in good agreement with
the values by Artemyev {\em et al.}~\cite{artemyev:05:pra}. We do not report results for $Z > 12$ for the
mixing states, since their uncertainty increases rapidly
(see Table~\ref{tab:ho:mix}) and the present method becomes less accurate than all-order calculations in
the $jj$ coupling \cite{artemyev:05:pra,malyshev:19,kozhedub:19}.

Table~\ref{tab:compar} compares our theoretical predictions for the intrashell $n = 2$ transition energies in helium-like
ions with results of previous calculations and available experimental data. We observe that for transitions between
non-mixing states, our calculation improves the theoretical accuracy typically by an order of magnitude as compared
to the previous calculations. Agreement with the experimental data is good and the theoretical results are more accurate
in the whole interval of $Z$ studied. By contrast, for transitions involving the mixing states the theoretical accuracy
is significantly lower.
Results for the fine-structure intervals are presented for the sake of completeness since
more accurate calculations with full inclusion of the $m\alpha^7$ effects are available in this case
\cite{pachucki:10:hefs}.

Summarizing, we performed a high-precision calculation of the two-electron QED effects to all orders in the
nuclear binding strength parameter $\Za$ and identified the higher-order QED effects of order
$m\alpha^7$ and higher. By using the ``unified'' approach, we combined together the NRQED calculation
of Ref.~\cite{yerokhin:10:helike} complete to order $m\alpha^6$ and all-order calculations of
one-electron and two-electron QED effects. In the result we obtained improved theoretical predictions
for ionization energies of the ground and non-mixing $n=2$ states of helium-like ions with $Z = 5\,$--$\,30$.
Theoretical predictions for the mixing $2\,^1 P_1$ and $2\,^3 P_1$ states are obtained for $Z = 5\,$--$\,12$.
Their accuracy is lower than that for the non-mixing states since the higher-order QED effects
are included within the one-electron approximation only. In order to advance the theory of the mixing states
further, one needs to extend the NRQED approach to embrace the perturbation theory of quasidegenerate states.

\begin{acknowledgments}
The authors are grateful to A.~N.~Artemyev for participation at the earlier stages of the calculations
and to A.~V.~Malyshev for useful discussions.
The work was supported by the Russian Science Foundation (Grant No. 20-62-46006).
K.P. acknowledges support from the National Science Center (Poland) Grant No. 2017/27/B/ST2/02459.
\end{acknowledgments}

\begin{table*}
\caption{Theoretical ionization energies of the ground and the non-mixing $n = 2$ excited states of helium-like ions,
in eV, $1\,{\rm eV} = 27.211\,386\,245\,988$~a.u.
\label{tab:IE:nonmix}}
\begin{ruledtabular}
\scriptsize
\begin{tabular}{llw{3.9}w{3.9}w{3.9}w{3.9}w{3.9}l}
\multicolumn{1}{l}{$Z$}
&\multicolumn{1}{l}{$A$}
& \multicolumn{1}{c}{$1\,^1 S_0$}
        & \multicolumn{1}{c}{$2\,^1 S_0$}
        & \multicolumn{1}{c}{$2\,^3 S_1$}
        & \multicolumn{1}{c}{$2\,^3 P_0$}
        & \multicolumn{1}{c}{$2\,^3 P_2$}
        & \multicolumn{1}{c}{Ref.}
  \\
\hline\\[-5pt]
%
%
%
 5 &11& 259.374\,4095\,(15)    & 56.569\,281\,05\,(80)  & 60.806\,181\,969\,(85) & 56.417\,932\,23\,(60)  & 56.413\,410\,95\,(26)   \\
   &  & 259.374\,40\,(2)       & 56.569\,277\,(15)      & 60.806\,181\,(17)      & 56.417\,930\,9\,(5)    & 56.413\,410\,4\,(4) & \cite{yerokhin:10:helike}\\
 6 &12& 392.090\,5875\,(26)    & 87.702\,7960\,(14)     & 93.128\,352\,06\,(18)  & 87.685\,5982\,(11)     & 87.670\,310\,62\,(50)   \\
 7 &14& 552.067\,4680\,(43)    & 125.652\,8520\,(24)    & 132.275\,521\,40\,(34) & 125.776\,2109\,(20)    & 125.739\,052\,56\,(88)  \\
 8 &16& 739.327\,0757\,(67)    & 170.427\,3032\,(39)    & 178.254\,625\,55\,(62) & 170.694\,2194\,(33)    & 170.618\,5383\,(15)   \\
 9 &19& 953.898\,447\,(10)     & 222.035\,5951\,(59)    & 231.075\,0281\,(11)    & 222.446\,8404\,(50)    & 222.309\,3293\,(23)   \\
10 &20& 1195.808\,475\,(15)    & 280.486\,8810\,(86)    & 290.746\,1290\,(17)    & 281.042\,6591\,(74)    & 280.812\,2429\,(34)   \\
11 &23& 1465.099\,543\,(22)    & 345.794\,144\,(12)     & 357.281\,4596\,(27)    & 346.492\,900\,(11)     & 346.129\,6103\,(49)   \\
12 &24& 1761.805\,457\,(29)    & 417.968\,719\,(16)     & 430.692\,9183\,(30)    & 418.809\,227\,(14)     & 418.263\,0745\,(64)   \\
   &  & 1761.804\,9\,(10)      & 417.968\,56\,(11)      & 430.692\,90\,(11)      & 418.809\,14\,(2)       & 418.263\,03\,(2)& \cite{yerokhin:10:helike}\\
   &  & 1761.804\,7\,(2)       & 417.968\,9\,(1)        & 430.692\,9\,(1)        & 418.809\,2\,(1)        & 418.263\,0\,(1) & \cite{artemyev:05:pra}\\
13 &27& 2085.977\,675\,(39)    & 497.026\,358\,(21)     & 510.996\,9945\,(43)    & 498.005\,959\,(18)     & 497.215\,7863\,(86)   \\
14 &28& 2437.658\,788\,(51)    & 582.981\,104\,(26)     & 598.208\,4225\,(55)    & 584.097\,729\,(23)     & 582.990\,066\,(11)    \\
15 &31& 2816.909\,423\,(63)    & 675.851\,568\,(33)     & 692.346\,6318\,(70)    & 677.101\,841\,(30)     & 675.589\,728\,(14)    \\
16 &32& 3223.781\,337\,(80)    & 775.654\,618\,(41)     & 793.429\,2528\,(89)    & 777.035\,871\,(37)     & 775.017\,688\,(17)    \\
17 &35& 3658.344\,41\,(10)     & 882.411\,785\,(51)     & 901.478\,679\,(12)     & 883.920\,107\,(46)     & 881.278\,372\,(22)    \\
18 &40& 4120.666\,36\,(13)     & 996.144\,406\,(62)     & 1016.517\,087\,(15)    & 997.775\,684\,(57)     & 994.375\,840\,(26)    \\
   &  & 4120.667\,2\,(9)       & 996.144\,3\,(3)        & 1016.516\,8\,(2)       & 997.775\,4\,(2)        & 994.375\,6\,(2)    & \cite{malyshev:19} \\
   &  & 4120.665\,3\,(4)       & 996.144\,6\,(2)        & 1016.517\,0\,(1)       & 997.775\,8\,(3)        & 994.375\,7\,(1)    & \cite{artemyev:05:pra}\\
19 &39& 4610.807\,89\,(16)     & 1116.872\,475\,(75)    & 1138.565\,273\,(19)    & 1118.624\,164\,(69)    & 1114.313\,372\,(32)   \\
20 &40& 5128.858\,40\,(19)     & 1244.623\,304\,(89)    & 1267.651\,485\,(24)    & 1246.490\,895\,(82)    & 1241.096\,747\,(38)   \\
21 &45& 5674.904\,20\,(23)     & 1379.423\,80\,(11)     & 1403.803\,533\,(30)    & 1381.402\,010\,(98)    & 1374.731\,277\,(45)   \\
22 &48& 6249.023\,07\,(28)     & 1521.299\,10\,(12)     & 1547.047\,425\,(37)    & 1523.383\,92\,(12)     & 1515.221\,274\,(53)   \\
23 &51& 6851.311\,46\,(34)     & 1670.279\,22\,(15)     & 1697.414\,119\,(45)    & 1672.465\,89\,(14)     & 1662.572\,575\,(62)   \\
24 &52& 7481.863\,27\,(41)     & 1826.394\,10\,(17)     & 1854.934\,471\,(54)    & 1828.678\,23\,(16)     & 1816.790\,721\,(72)   \\
25 &55& 8140.787\,72\,(49)     & 1989.677\,64\,(20)     & 2019.643\,353\,(66)    & 1992.053\,73\,(18)     & 1977.882\,359\,(84)   \\
26 &56& 8828.188\,09\,(58)     & 2160.162\,56\,(23)     & 2191.574\,459\,(79)    & 2162.625\,73\,(21)     & 2145.853\,312\,(96)   \\
   &  & 8828.189\,6\,(25)      & 2160.162\,5\,(8)       & 2191.574\,2\,(7)       & 2162.625\,3\,(7)       & 2145.853\,0\,(7)   & \cite{kozhedub:19}\\
   &  & 8828.187\,5\,(11)      & 2160.163\,2\,(7)       & 2191.574\,5\,(6)       & 2162.626\,1\,(10)      & 2145.853\,2\,(2)   & \cite{artemyev:05:pra}\\
27 &59& 9544.183\,39\,(68)     & 2337.886\,07\,(26)     & 2370.765\,981\,(94)    & 2340.430\,43\,(24)     & 2320.710\,82\,(11)    \\
28 &58& 10288.886\,21\,(80)    & 2522.883\,72\,(29)     & 2557.254\,48\,(11)     & 2525.504\,31\,(28)     & 2502.460\,95\,(13)    \\
29 &63& 11062.431\,11\,(94)    & 2715.197\,98\,(34)     & 2751.083\,47\,(13)     & 2717.888\,06\,(32)     & 2691.112\,45\,(14)    \\
30 &64& 11864.939\,4\,(11)     & 2914.866\,90\,(38)     & 2952.292\,00\,(15)     & 2917.621\,11\,(36)     & 2886.671\,37\,(16)    \\
\end{tabular}
\end{ruledtabular}
\end{table*}

\begin{table}
\caption{Same as Table~\ref{tab:IE:nonmix} but for the mixing $n =2$ states.
\label{tab:IE:mix}}
\begin{ruledtabular}
\begin{tabular}{lw{3.9}w{3.9}l}
\multicolumn{1}{l}{$Z$}
        & \multicolumn{1}{c}{$2\,^1 P_1$}
        & \multicolumn{1}{c}{$2\,^3 P_1$}
        & \multicolumn{1}{c}{Ref.}
  \\
\hline\\[-5pt]
%
 5 & 53.809\,194\,15\,(36)  & 56.419\,9395\,(14)     \\
   & 53.809\,194\,4\,(1)    & 56.419\,9393\,(4) & \cite{yerokhin:10:helike} \\
 6 & 84.188\,092\,04\,(92)    & 87.687\,1467\,(31)     \\
 7 & 121.371\,7397\,(23)    & 125.775\,1315\,(64)    \\
 8 & 165.365\,9011\,(55)    & 170.686\,923\,(12)     \\
 9 & 216.176\,651\,(13)     & 222.428\,084\,(22)     \\
10 & 273.807\,646\,(28)     & 281.005\,369\,(41)     \\
11 & 338.266\,548\,(59)     & 346.428\,077\,(75)     \\
12 & 409.556\,49\,(12)      & 418.705\,94\,(13)      \\
   & 409.556\,460\,(6)      & 418.705\,96\,(17) & \cite{yerokhin:10:helike} \\
   & 409.556\,4\,(1)        & 418.705\,9\,(1) & \cite{artemyev:05:pra} \\
\end{tabular}
\end{ruledtabular}
\end{table}


\begin{thebibliography}{10}

\bibitem{indelicato:19}
P.~Indelicato,
\newblock J. Phys. B {\bf 52}, 232001 (2019).

\bibitem{artemyev:05:pra}
A.~N. Artemyev, V.~M. Shabaev, V.~A. Yerokhin, G.~Plunien, and G.~Soff,
\newblock Phys. Rev. A {\bf 71}, 062104 (2005).

\bibitem{kozhedub:19}
Y.~S. Kozhedub, A.~V. Malyshev, D.~A. Glazov, V.~M. Shabaev, and I.~I.
  Tupitsyn,
\newblock Phys. Rev. A {\bf 100}, 062506 (2019).

\bibitem{malyshev:19}
A.~V. Malyshev, Y.~S. Kozhedub, D.~A. Glazov, I.~I. Tupitsyn, and V.~M.
  Shabaev,
\newblock Phys. Rev. A {\bf 99}, 010501 (2019).

\bibitem{pachucki:17:heSummary}
K.~Pachucki, V.~Patk\'o\v{s}, and V.~A. Yerokhin,
\newblock Phys. Rev. A {\bf 95}, 062510 (2017).

\bibitem{yerokhin:21:hereview}
V.~A. Yerokhin, V.~Patk\'o\v{s}, and K.~Pachucki,
\newblock Symmetry {\bf 13}, 1246 (2021).

\bibitem{drake:88:cjp}
G.~W.~F. Drake,
\newblock Can. J. Phys. {\bf 66}, 586  (1988).

\bibitem{yerokhin:10:helike}
V.~A. Yerokhin and K.~Pachucki,
\newblock Phys. Rev. A {\bf 81}, 022507 (2010).

\bibitem{yerokhin:20:fs}
V.~A. Yerokhin, M.~Puchalski, and K.~Pachucki,
\newblock Phys. Rev. A {\bf 102}, 042816 (2020).

\bibitem{patkos:21:helamb}
V.~Patk\'o\v{s}, V.~A. Yerokhin, and K.~Pachucki,
\newblock Phys. Rev. A {\bf 103}, 042809 (2021).

\bibitem{clausen:21}
G.~Clausen, P.~Jansen, S.~Scheidegger, J.~A. Agner, H.~Schmutz, and F.~Merkt,
\newblock Phys. Rev. Lett. {\bf 127}, 093001 (2021).

\bibitem{yerokhin:18:betherel}
V.~A. Yerokhin, V.~Patk\'o\v{s}, and K.~Pachucki,
\newblock Phys. Rev. A {\bf 98}, 032503 (2018),
\newblock {\em ibid.} {\bf 103}, 029901(E) (2021).

\bibitem{patkos:20}
V.~Patk\'o\v{s}, V.~A. Yerokhin, and K.~Pachucki,
\newblock Phys. Rev. A {\bf 101}, 062516 (2020),
\newblock {\em ibid.} {\bf 103}, 029902(E) (2021).

\bibitem{yerokhin:97:pla}
V.~A. Yerokhin, A.~N. Artemyev, and V.~M. Shabaev,
\newblock Phys. Lett. A {\bf 234}, 361  (1997).

\bibitem{yerokhin:20:green}
V.~A. Yerokhin and A.~V. Maiorova,
\newblock Symmetry {\bf 12}, 800 (2020).

\bibitem{yerokhin:99:pra}
V.~A. Yerokhin and V.~M. Shabaev,
\newblock Phys. Rev. A {\bf 60}, 800  (1999).

\bibitem{yerokhin:05:se}
V.~A. Yerokhin, K.~Pachucki, and V.~M. Shabaev,
\newblock Phys. Rev. A {\bf 72}, 042502 (2005).

\bibitem{yerokhin:99:sescr}
V.~A. Yerokhin, A.~N. Artemyev, T.~Beier, G.~Plunien, V.~M. Shabaev, and
  G.~Soff,
\newblock Phys. Rev. A {\bf 60}, 3522  (1999).

\bibitem{yerokhin:20:gfact}
V.~A. Yerokhin, K.~Pachucki, M.~Puchalski, C.~H. Keitel, and Z.~Harman,
\newblock Phys. Rev. A {\bf 102}, 022815 (2020).

\bibitem{persson:96:2el}
H.~Persson, S.~Salomonson, P.~Sunnergren, and I.~Lindgren,
\newblock Phys. Rev. Lett. {\bf 76}, 204 (1996).

\bibitem{artemyev:97}
A.~N. Artemyev, V.~M. Shabaev, and V.~A. Yerokhin,
\newblock Phys. Rev. A {\bf 56}, 3529 (1997).

\bibitem{fainshtein:91}
A.~G. Fainshtein, N.~L. Manakov, and A.~A. Nekipelov,
\newblock J. Phys. B {\bf 24}, 559  (1991).

\bibitem{manakov:12:vgu}
N.~L. Manakov and A.~A. Nekipelov,
\newblock Vestnik Voronezhskogo gosudarstvennogo universiteta {\bf 2}, 53
  (2012),
\newblock [in Russian],
  http://www.vestnik.vsu.ru/pdf/physmath/2012/02/2012-02-07.pdf.

\bibitem{manakov:13:vgu}
N.~L. Manakov and A.~A. Nekipelov,
\newblock Vestnik Voronezhskogo gosudarstvennogo universiteta {\bf 2}, 84 (2013),
\newblock [in Russian],
  http://www.vestnik.vsu.ru/pdf/physmath/2013/02/2013-02-08.pdf.

\bibitem{johnson:88}
W.~R. Johnson, S.~A. Blundell, and J.~Sapirstein,
\newblock Phys. Rev. A {\bf 37}, 307  (1988).

\bibitem{artemyev:20:priv}
A.~N. Artemyev,
\newblock Private communication, 2021.

\bibitem{blundell:93:b}
S.~A. Blundell, P.~J. Mohr, W.~R. Johnson, and J.~Sapirstein,
\newblock Phys. Rev. A {\bf 48}, 2615  (1993).

\bibitem{mohr:00:pra}
P.~J. Mohr and J.~Sapirstein,
\newblock Phys. Rev. A {\bf 62}, 052501 (2000).

\bibitem{andreev:01}
O.~Y. Andreev, L.~N. Labzowsky, G.~Plunien, and G.~Soff,
\newblock Phys. Rev. A {\bf 64}, 042513 (2001).

\bibitem{asen:02}
B.~\r{A}sen, S.~Salomonson, and I.~Lindgren,
\newblock Phys. Rev. A {\bf 65}, 032516 (2002).

\bibitem{andreev:04}
O.~Y. Andreev, L.~N. Labzowsky, G.~Plunien, and G.~Soff,
\newblock Phys. Rev. A {\bf 69}, 062505 (2004).

\bibitem{shabaev:02:rep}
V.~M. Shabaev,
\newblock Phys. Rep. {\bf 356}, 119  (2002).

\bibitem{lindgren:95:pra}
I.~Lindgren, H.~Persson, S.~Salomonson, and L.~Labzowsky,
\newblock Phys. Rev. A {\bf 51}, 1167  (1995).

\bibitem{shabaev:04:DKB}
V.~M. Shabaev, I.~I. Tupitsyn, V.~A. Yerokhin, G.~Plunien, and G.~Soff,
\newblock Phys. Rev. Lett. {\bf 93}, 130405 (2004).

\bibitem{yerokhin:21:gfact}
V.~A. Yerokhin, C.~H. Keitel, and Z.~Harman,
\newblock Phys. Rev. A {\bf 104}, 022814 (2021).

\bibitem{yerokhin:18:hydr}
V.~A. Yerokhin, K.~Pachucki, and V.~Patk\'o\v{s},
\newblock Ann. Phys. (Leipzig) {\bf 531}, 1800324 (2019).

\bibitem{karshenboim:97}
S.~G. Karshenboim,
\newblock Z. Phys. D {\bf 39}, 109 (1997).

\bibitem{mohr:75:prl}
P.~J. Mohr,
\newblock Phys. Rev. Lett. {\bf 34}, 1050  (1975).

\bibitem{yerokhin:15:Hlike}
V.~A. Yerokhin and V.~M. Shabaev,
\newblock J. Phys. Chem. Ref. Data {\bf 44}, 033103 (2015).

\bibitem{drake:82:nimb}
G.~W.~F. Drake,
\newblock Nucl. Instrum. Methods {\bf B 202}, 273  (1982).

\bibitem{malyshev:21:belike}
A.~Malyshev, Y.~Kozhedub, I.~Anisimova, D.~Glazov, M.~Kaygorodov, I.~Tupitsyn,
  and V.~Shabaev,
\newblock Optics and Spectroscopy {\bf 129}, 652 (2021).

\bibitem{wang:12}
M.~Wang, G.~Audi, A.~H. Wapstra, F.~G. Kondev, M.~MacCormick, X.~Xu, and
  B.~Pfeiffer,
\newblock Chin. Phys. C {\bf 36}, 1603  (2012).

\bibitem{angeli:13}
I.~Angeli and K.~Marinova,
\newblock At. Dat. Nucl. Dat. Tabl. {\bf 99}, 69  (2013).

\bibitem{thompson:98}
J.~K. Thompson, D.~J.~H. Howie, and E.~G. Myers,
\newblock Phys. Rev. A {\bf 57}, 180 (1998).

\bibitem{myers:99:Mg}
E.~G. Myers and M.~R. Tarbutt,
\newblock Phys. Rev. A {\bf 61}, 010501 (1999).

\bibitem{myers:96}
E.~G. Myers, D.~J.~H. Howie, J.~K. Thompson, and J.~D. Silver,
\newblock Phys. Rev. Lett. {\bf 76}, 4899  (1996).

\bibitem{dinneen:91}
T.~P. Dinneen, N.~Berrah-Mansour, H.~G. Berry, L.~Young, and R.~C. Pardo,
\newblock Phys. Rev. Lett. {\bf 66}, 2859 (1991).

\bibitem{peacock:84}
N.~J. Peacock, M.~F. Stamp, and J.~D. Silver,
\newblock Phys. Scr. {\bf T8}, 10 (1984).

\bibitem{curdt:00}
W.~Curdt, E.~Landi, K.~Wilhelm, and U.~Feldman,
\newblock Phys. Rev. A {\bf 62}, 022502 (2000).

\bibitem{kukla:95}
K.~W. Kukla, A.~E. Livingston, J.~Suleiman, H.~G. Berry, R.~W. Dunford, D.~S.
  Gemmell, E.~P. Kanter, S.~Cheng, and L.~J. Curtis,
\newblock Phys. Rev. A {\bf 51}, 1905 (1995).

\bibitem{buchet:81}
J.~P. Buchet, M.~C. Buchet-Poulizac, A.~Denis, J.~D\'esesquelles, M.~Druetta,
  J.~P. Grandin, and X.~Husson,
\newblock Phys. Rev. A {\bf 23}, 3354 (1981).

\bibitem{klein:85}
H.~A. Klein, F.~Moscatelli, E.~G. Myers, E.~H. Pinnington, J.~D. Silver, and
  E.~Trabert,
\newblock J. Phys. B {\bf 18}, 1483 (1985).

\bibitem{pachucki:10:hefs}
K.~Pachucki and V.~A.~Yerokhin,
\newblock Phys.~Rev.~Lett.~ {\bf 104}, 070403 (2010)

\end{thebibliography}

\begin{table}
\caption{Comparison of theoretical and experimental $n = 2$ intrashell transition energies, in cm$^{-1}$.
\label{tab:compar}}
\begin{ruledtabular}
\begin{tabular}{lw{2.9}w{5.9}w{1.5}l}
\multicolumn{1}{l}{$Z$}
        & \multicolumn{1}{c}{Theory}
        & \multicolumn{1}{c}{Experiment}
        & \multicolumn{1}{c}{Difference}
        & \multicolumn{1}{c}{Ref.}
  \\
\hline\\[-5pt]
$2\,^3 S_1$--$2\,^3 P_0$ \\
 5 & 35\,393.6211\,(49) & 35\,393.627\,(13) & -0.006\,(13)  & \cite{dinneen:91} \\
   & 35\,393.628\,(14)^a\\[2pt]
 8 & 60\,978.788\,(27)  & 60\,978.44\,(52)  &  0.35\,(52)   &\cite{peacock:84} \\
   & 60\,978.85\,(14)^a \\[2pt]
 12& 95\,848.43\,(11)   & 95\,851.27\,(92)  &  0.15\,(93)   & \cite{curdt:00} \\
   & 95\,849.0\,(9)^a   \\
   & 95\,848.(1)^b   \\[2pt]
 14& 113\,810.42\,(19)  & 113\,806.7\,(3.7) &  3.7\,(3.7)   & \cite{curdt:00} \\
   & 113\,809.(2)^b  \\[2pt]
 18& 151\,159.61\,(47)  & 151\,164.0\,(4.1) & -4.4\,(4.1)   & \cite{kukla:95} \\
   & 151\,158.(3)^b  \\[2pt]
 26& 233\,487.3(1.8)    & 232\,558.\,(550)  & 71.(550)      & \cite{drake:88:cjp}\\
   & 233\,484.(10)^b  \\
   & 233\,485.(9)^c   \\[3pt]
$2\,^3 S_1$--$2\,^3 P_2$ \\
 5 & 35\,430.0876\,(22) & 35\,430.084\,(9)  &   0.004\,(9)  & \cite{dinneen:91} \\
   & 35\,430.088\,(14)^a \\[2pt]
 7 & 52\,720.1766\,(76) & 52\,720.23\,(69)  &  -0.05\,(69)  & \cite{peacock:84} \\
   & 52\,720.18\,(7)^a \\[2pt]
 8 & 61\,589.198\,(13)  & 61\,589.70\,(53)  &  -0.50\,(53)  & \cite{peacock:84} \\
   & 61\,589.21\,(14)^a \\[2pt]
 10& 80\,122.195\,(31)  & 80\,123.33\,(83)  &  -1.1\,(0.8)  & \cite{peacock:84} \\
   & 80\,122.3\,(4)^a   \\[2pt]
 12& 100\,253.451\,(57) & 100\,255.9\,(1.9) &  -2.5\,(1.9)  & \cite{curdt:00}   \\
   & 100\,253.7\,(9)^a \\
   & 100\,253.\,(1)^b  \\[2pt]
 14& 122\,744.326\,(98) & 122\,740.4\,(3.6) &   3.9\,(3.6)  &\cite{curdt:00}    \\
   & 122\,744.(1)^b \\[2pt]
 18& 178\,581.19\,(24)  & 178\,589.3\,(5.1) &  -8.1\,(5.1)  & \cite{kukla:95}   \\
   & 178\,581.(2)^b \\[2pt]
 26& 368\,765.9(1.0)    & 368\,976.(125)    & -210.(125)    & \cite{buchet:81} \\
   & 368\,767.(6)^b \\
   & 368\,767.(5)^c \\[3pt]
$2\,^3 S_1$--$2\,^3 P_1$ \\
 5 & 35\,377.432\,(11)  & 35\,377.424\,(13) &  0.008\,(17)  & \cite{dinneen:91} \\
   & 35\,377.429\,(14)^a&                    \\[2pt]
 8 & 61\,037.634\,(99)  & 61\,037.62\,(93)  &  0.01\,(93)   & \cite{peacock:84} \\
   & 61\,037.65\,(14)^a &                    \\[2pt]
 12& 966\,81.5\,(1.1)   & 966\,83.(6)       & -2.\,(6)      & \cite{klein:85} \\
   & 966\,82.(1)^a \\[3pt]
$2\,^1 S_0$--$2\,^3 P_1$ \\
 7 &  986.251\,(55)     & 986.3180\,(7)     &  -0.07\,(6)   & \cite{thompson:98}  \\
   &  986.36\,(7)^a \\[3pt]
$2\,^3 P_0$--$2\,^3 P_1$ \\
 7 &  8.706\,(54)       & 8.6707\,(7)        &   0.035\,(54)& \cite{thompson:98} \\
   &  8.675\,(21)^a     \\
   &  8.6731\,(67)^d    \\[2pt]
 12&  833.0\,(1.1)      & 833.133\,(15)      &  -0.1\,(1.1) & \cite{myers:99:Mg} \\
   &  832.2\,(0.2)^a   \\
   &  834.(1)^b        \\[3pt]
$2\,^3 P_2$--$2\,^3 P_1$ \\
 9 &  957.82\,(18)      & 957.8730\,(12)     &  -0.05\,(18) & \cite{myers:96} \\
   &  957.797\,(54)^a \\
   &  957.886\,(79)^d
\end{tabular}
\end{ruledtabular}
$^a$~Yerokhin and Pachucki 2010 \cite{yerokhin:10:helike};
$^b$~Artemyev {\em et al.} 2005 \cite{artemyev:05:pra};
$^c$~Kozhedub {\em et al.} 2019 \cite{kozhedub:19};
$^d$~Pachucki and Yerokhin 2010 \cite{pachucki:10:hefs}.\\
\end{table}

\end{document}